\documentclass[12pt]{article}

\RequirePackage{amsmath,amsfonts,amssymb}
\usepackage{multirow}
\usepackage{hhline}
\usepackage{subfigure}
\usepackage[utf8]{inputenc}
\usepackage{url}
\usepackage{graphicx}
\usepackage{times}
\usepackage{cprotect}
\usepackage{lettrine}
\usepackage{textcomp}
\usepackage[dvipsnames]{xcolor}
\usepackage{lineno, blindtext}

\newcommand{\matr}[1]{\mathbf{#1}}

\usepackage[labelsep=period]{caption}
\usepackage[labelfont=bf]{caption}

\title{Shifting Polarization and Twitter News Influencers between two U.S. Presidential Elections}

\author
{
James Flamino$^{\text{1}}$, Alessandro Galezzi$^{\text{2}}$, Stuart Feldman$^{\text{3}}$, Michael W. Macy$^{\text{4}}$,\\
Brendan Cross$^{\text{1}}$, Zhenkun Zhou$^{\text{5}}$, Matteo Serafino$^{\text{6}}$, Alexandre Bovet$^{\text{7}}$,\\
 Hern\'an A. Makse$^{\text{8}}$, Boleslaw K. Szymanski$^{\text{1,9,*}}$\\
\normalsize{$^{\text{1}}$Department of Computer Science and Network Science and Technology Center,}\\
\normalsize{Rensselaer Polytechnic Institute, Troy, NY, USA}\\
\normalsize{$^{\text{2}}$University of Brescia, Brescia, Italy}\\
\normalsize{$^{\text{3}}$Schmidt Futures, New York, NY, USA}\\
\normalsize{$^{\text{4}}$Departments of Information Science and Sociology, Cornell University, Ithaca, NY, USA}\\
\normalsize{$^{\text{5}}$School of Statistics, Capital University of Economics and Business, Beijing, China}\\
\normalsize{$^{\text{6}}$IMT School for Advanced Studies, 55100 Lucca, Italy}\\
\normalsize{$^{\text{7}}$Mathematical Institute, University of Oxford, UK}\\
\normalsize{$^{\text{8}}$Levich Institute and Physics Department, City College of New York, New York, NY, USA}\\
\normalsize{$^{\text{9}}$Spo\l{}eczna Akademia Nauk, \L{}\'{o}d\'{z}, Poland}\\
\normalsize{$^{*}$Corresponding author, Email: szymab@rpi.edu}
}

\date{}

\begin{document} 

\maketitle 


\begin{abstract}
Social media are decentralized, interactive, and transformative, empowering users to produce and spread information to influence others. This has changed the dynamics of political communication that were previously dominated by traditional corporate news media. Having hundreds of millions of tweets collected over the 2016 and 2020 U.S. presidential elections gave us a unique opportunity to measure the change in polarization and the diffusion of political information. We analyze the diffusion of political information among Twitter users and investigate the change of polarization between these elections and how this change affected the composition and polarization of influencers and their retweeters. We identify ``influencers'' by their ability to spread information and classify them into those affiliated with a media organization, a political organization, or unaffiliated. Most of the top influencers were affiliated with media organizations during both elections. We found a clear increase from 2016 to 2020 in polarization among influencers and among those whom they influence. Moreover, 75\% of the top influencers in 2020 were not present in 2016, demonstrating that such status is difficult to retain. Between 2016 and 2020, 10\% of influencers affiliated with media were replaced by center- or right-orientated influencers affiliated with political organizations and unaffiliated influencers. 
\end{abstract}

\section*{Introduction}

A growing number of studies have documented increasing political polarization in the U.S. that is deeper than at any time since the American Civil War~\cite{brady2006polarization,hare2014polarization,axelrod2021preventing}. Partisan division over issues has increased among those affiliated with political and news media organizations -- elected representatives, party officials, and political pundits -- alongside an alarming increase in affective polarization among voters. This two-level pattern -- issue polarization among political elites and affective polarization among voters -- invites further research on the diffusion of polarized political information between those in positions of political influence and the larger population.

This diffusion of political information is difficult to track with traditional survey and roll call voting data that lack relational measures. Increasing reliance on social media for political communication is opening up unprecedented opportunities to study the diffusion of political information and misinformation~\cite{vosoughi2018spread} over communication networks~\cite{guilbeault2021topological}. Our study leverages social media data from Twitter to better understand the diffusion dynamics of news media information during the two most recent U.S. election cycles.  

Twitter users are embedded in relatively stable communication networks created by the exchange of ``retweets.'' A 2015 study by Metaxas et al.~\cite{metaxas2015retweets} found that ``retweeting indicates not only interest in a message, but also trust in the message and the originator, and agreement with the message contents.'' The content of retweets makes it possible to identify information that is highly biased, as well as the ideological direction of the bias. Using retweet data we also can identify ``influencers'' who are the users with the greatest ability to broadly propagate new information over the retweet network. Typically, influencer tweets are highly likely to be retweeted, not only by their followers, but also by their followers' followers, and so on. We classify Twitter influencers into two categories. The first includes the ``affiliated'' who are associated with media or political organizations, and the other consists of the ``unaffiliated'' who do not have such associations, so most likely represent themselves or informal groups.

Our study also aims to better understand how polarization unfolds on social media. To clarify, political scientists distinguish multiple types and levels of polarization \cite{mccarty2019polarization,galston2006delineating,abramowitz2013polarized,fiorina2008political,layman2006party,mason2015disrespectfully}: policy polarization (extreme differences of opinion on highly salient issues), partisan polarization (alignment of opinions with opposing political parties), ideological polarization (alignment of opinions with liberal vs. conservative world views), and geographic polarization (regional alignment of opinions, e.g., ``red state/blue state''). Each of these four types of polarization can in turn be classified by level: elite polarization among political officials and pundits, media polarization among news organizations, and polarization among the underlying population as usually measured by exit polls and opinion surveys.
In this paper, we use data from social media to study ideological polarization among the political elite, news organizations, and Twitter users more broadly. Over the past decade, the rapid growth of Twitter, Facebook, Reddit and other social media have transformed the communications and information propagation landscape. Alongside traditional broadcast media and face-to-face communication, people now have the ability to search for and exchange information with billions of other users in a global network. Recent studies have examined the impact of new technologies like Twitter and YouTube on election outcomes~\cite{effing2011social,broersma2012social,metaxas2012social,ceron2016politics,Bovet2018,soares2018influencers,grover2019,lee2019social,acharoui2020,suau-gomila2020}, including the effects of disinformation~\cite{Allcott2017,shao2018anatomy,Bovet2019,Grinberg2019,Ruths2019,machado2019study}. Other studies have documented how social media platforms contribute to polarization through the creation of echo chambers~\cite{Conover2011,prior2013media,Mocanu2015,Barbera2015b,Bessi2016,Vaccari2016,Bessi2016b,lelkes2017hostile,bail2018exposure,Cinelli2021}. In contrast, our study focuses on the diffusion of news media information between influencers and those whom they influence, as well as the change in composition, popularity, and polarization among influencers and their retweeters in the months leading up to the 2016 to 2020 U.S. presidential elections.

To maintain the consistency between the results from 2016 and 2020 elections, we follow the methodology of Ref.~\cite{Bovet2019} to identify and classify the influences and their retweeters in the 2020 U.S. election data.
We classify tweets containing a link to a news outlet into several news media categories corresponding to their different political orientations.
We observe that the volume of tweets and users with a center orientation decreased from the 2016 election to the 2020 election. 
For each news media category, we reconstruct the corresponding retweet network and identify the most important news media influencers of the category by finding the most important nodes in terms of their ability to spread information in the network. The top 25 influencers in each news media category are then classified as affiliated with a media or with a political organization or unaffiliated. 
We find that the proportion of top influencers affiliated with news media organizations decreased in 2020, while the proportion who were politically linked increased. This may indicate a shift in the sources of influence over political agenda setting.
Simultaneously, the proportion of media affiliated influencers increased in 2020 in the categories containing disinformation, indicating a shift in the sources of disinformation from informal to formal organizations.
Finally, we measure the strength of the polarization of the influencers and of their retweeters, defined as the level of separation of the influencers' retweeters in two opposite clusters and find a clear, significant increase of the polarization from 2016 to 2020.

\section*{Results}

\subsection*{News media on Twitter in 2016 and 2020}
We tracked the spread of political news on Twitter in 2016 and 2020 by analyzing two datasets containing tweets posted between June $1^{\text{st}}$ and election day (November $8^{\text{th}}$ in 2016 and November $2^{\text{nd}}$ in 2020). The data were collected continuously using the Twitter Search API with the names of the two presidential candidates in each of the presidential elections in 2016 and 2020 as keywords.
(Had we used more keywords targeting specific media outlets or hashtags concerning specific news events we would risk missing election-related tweets that did not contain references to the list of outlets or events.) The 2016 dataset contains 171 million tweets sent by 11 million users and was used in Refs.~\cite{Bovet2018,Bovet2019} to assess the influence of disinformation on Twitter in 2016. The 2020 dataset contains 702 million tweets sent by 20 million users. Hence, we observe a significant increase in Twitter involvement in distributing election polarization, since in four years the number of Twitter users nearly double and number of tweets per user more than double, increasing the total number of tweets more than fourfold. 

The classifications of news media websites presented below and used in this paper, including ``fake'', ``extremely biased'', ``left'', ``right'', and especially the boundaries between categories, are a matter of opinion, rather than a statement of fact. The categorizations and labels assigned to the corresponding classes used here originated in publicly available datasets from fact-checking and bias rating organizations credited below. The political views and conclusions contained in this article should not be interpreted as representing those of the authors or their funders.

For each tweet containing a URL link, we extracted the domain name of the URL (e.g. \texttt{www.cnn.com}) and classified each link directing to a news media outlet according to this outlet's political bias. The 2016 and 2020 classifications rely on the website \url{allsides.com} (AS), followed by the bias classification from \url{mediabiasfactcheck.com} (MBFC) for outlets not present in AS (both taken as of January 7 2021 for the 2020 classification). We classified URL links in five news media categories for outlets that mostly conform to professional standards of fact-based journalism: \textit{right}, \textit{right-leaning}, \textit{center}, \textit{left-leaning} and \textit{left}. We also include three additional news media categories to include outlets that tend to disseminate disinformation: \textit{extreme-right bias}, \textit{extreme-left bias} and \textit{fake news}. Websites in the \textit{fake news} category have been flagged by fact-checking organizations as spreading fabricated news or conspiracy theories, while websites in the extremely biased category have been flagged for reporting controversial information that distorts facts and may rely on propaganda, decontextualized information, or opinions misrepresented as facts. A detailed explanation of the methodologies used by AS and MBFC for rating news outlets and of the differences in classification between 2016 and 2020 is given in the Methods section. The full lists of outlets in each category in 2016 and 2020 are given in SM Tabs. S1 and S2. In the 2016 dataset, 30.7 million tweets, sent by 2.3 million users, contain a URL directed to a media outlet website. The 2020 dataset contained 72.7 million tweets with news links sent by 3.7 million users. This number reveals remarkable drop of fraction of flow of tweets from users associated with media form $18\%$ in 2016 to $10\%$ so nearly half lower. This came from mainly from smaller growth of productivity of media affiliated users.

\begin{figure*}[t!]
\centering
\includegraphics[width=0.9\linewidth]{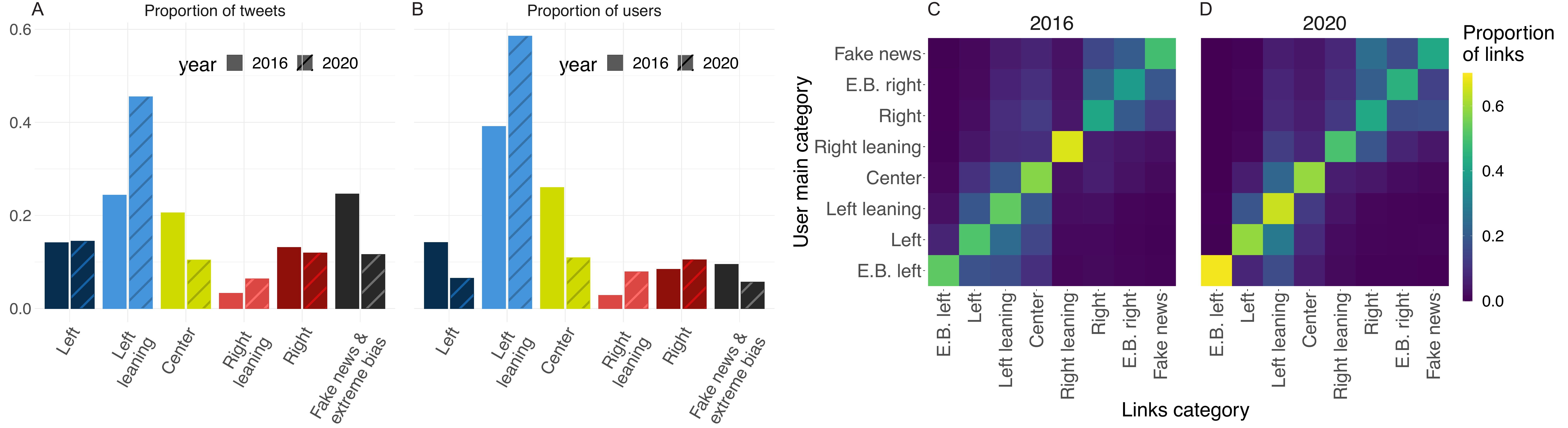}
\caption{\textbf{Distribution of news media links in 2016 and 2020, by news media category.} Panels A and B show the fractions of tweets and users that  sent tweets with a URL pointing to a website belonging to one of the categories.
Users are classified in the category in which they posted the most links.
For the users that have at least two links classified, panels C and D report the fraction of links across categories as a function of the users' main categories.}
\label{fig:num_tweets_num_users}
\end{figure*}

The fractions of tweets and users who sent a tweet in each of the news media categories are shown in Fig.~\ref{fig:num_tweets_num_users} A and B (the numbers are reported in SM Tab. S3) along with other statistics about the activity of users in each category. Between 2016 and 2020, these fractions decreased most in the center category and increased most in the left-leaning category, with a smaller increase in the fractions in the right-leaning category. The shift away from the center may indicate the increasing  polarization, both among  users as well as media outlets. However, most of the decrease in the fraction of center media outlets reflects the shift of \url{CNN.com}, which was categorized by AS as center in 2016 and as left-leaning in 2020, combined with \textit{CNN} accumulating more than twice the number of tweets in 2020 than the top outlet of the center category (\url{thehill.com}) (SM Tab. S2).

The fraction of tweets in the fake and extremely biased category, representing outlets that were most susceptible to sharing disinformation, decreased from 10\% to 6\% for fake news and from 13\% to 6\% for extremely right-bias news. The number of users who shared those tweets also decreased for extremely right-biased news (from 6\% to 3\%) but not for fake news (which remained at 3\%) (SM Tab. S3). The fraction of tweets in the extremely-left bias category is very small (2\% in 2016 and even less, 0.05\% in 2020).

Fig.~\ref{fig:num_tweets_num_users} C and D shows the fraction of URLs across categories as a function of a user's modal category for users that posted at least two links in our datasets. 
The analysis reveals two clusters in 2016 and 2020, one with categories from the right and fake news (fake news, extreme-right bias news, right news and right-leaning news) and the second one with categories from the center and left (center news, left-leaning news, left news, extreme-left bias news). These two clusters can be interpreted as two echo chambers. Asymmetrical patterns in Fig.~\ref{fig:num_tweets_num_users} C and D reveal that users in the right wing echo chamber also link to a very limited number of left wing media outlets, but that the opposite relation does not occur. This is consistent with asymmetry between left-leaning and right-leaning users in social media observed in previous studies~\cite{Cinelli2021,Bovet2019,Bakshy2015}.

In order to estimate the volume of tweets sent from automated accounts such as bots, we counted the number of tweets sent from unofficial Twitter clients, e.g., Twitter clients other than the Twitter Web client, Android client, IPhone client or other official clients. 
Unofficial Twitter clients include a variety of different applications used to automate all or part of an account activity, such as third party applications used typically by brands and professionals (e.g. SocialFlow or Hootsuite) or bots created with malicious intentions~\cite{Bovet2018}.

The overall fraction of tweets sent from unofficial clients was 8\% in 2016 and 1\% in 2020. A similar drop over the same period was observed in the average activity of these users (see SM Tab. S3). This decrease could be attributed in part to measures taken by Twitter to limit the virality of disinformation. Our results show that while the relative volume of tweets linking to disinformation websites dropped approximately by a half in 2020 compared to 2016, the fraction of users sharing fake news decreased significantly (Fig.~\ref{fig:num_tweets_num_users} A and B and Tab. S3).

\begin{figure}[ht]
\centering
\includegraphics[width=0.7\linewidth]{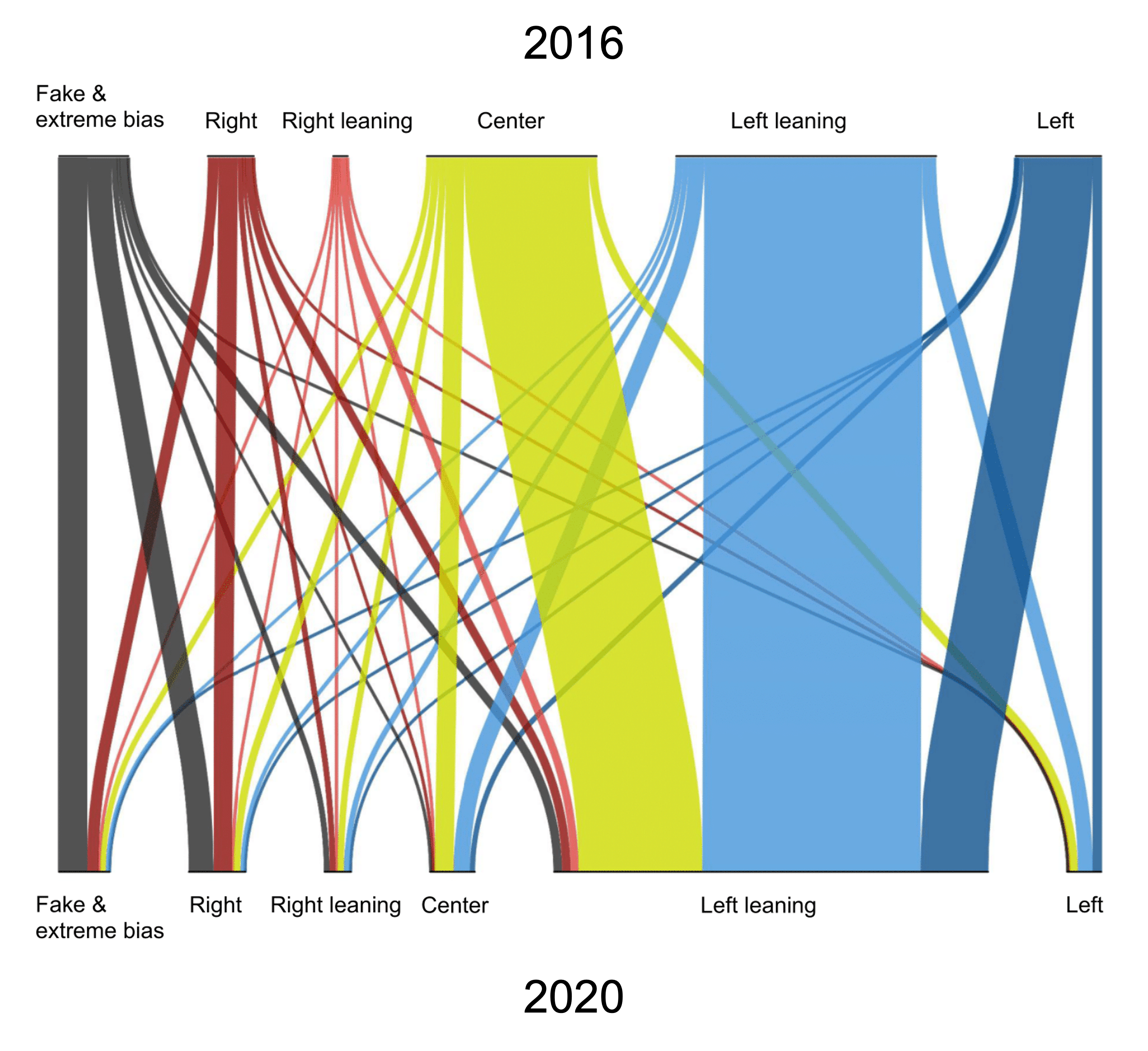}
\caption{\textbf{Shifts of users across news media categories from 2016 to 2020.} The size of each category in 2016 corresponds to the number of unique users in the category in 2016 (Fig.~\ref{fig:num_tweets_num_users}). The shift from one category to another is proportional to the fraction of users that were classified in 2016 and in 2020 in the two respective categories.
}
\label{fig:user_flow}
\end{figure}

To understand how users shifted between categories from 2016 to 2020, we track users that are present in both election datasets and in both years are classified into the category in which they posted the most tweets in each year. Fig.~\ref{fig:user_flow} shows the resulting shifts. The two largest of them are of users that were in the center and left news category in 2016 and shifted to the left-leaning category in 2020. This reflects the consolidation of the left-leaning category as the largest in 2020, with the three most widely shared outlets: \textit{New York Times}, \textit{Washington Post} and \textit{CNN} (SM Tab. S2). We also observe a large fraction of users in the fake and extremely biased news category in 2016 that moved to the right news category in 2020. However, these user shifts also reflect the change in the classification of media outlets from 2016 to 2020. We infer the ideological position of Twitter users without relying on the news outlet classification in section \ref{sec:twitter_pola}, and show that the resulting positions are highly correlated with the user positions computed using the news categories in which they posted.




Instead, we use directed edges to track the flow of the information contained in the tweet from the node that originated the information to those who retweet it.


\subsection*{News media influencers}

To capture the dynamics of information diffusion, we reconstruct retweet networks corresponding to each news media category.
We add a link, or directed edge, going from node $v$ to node $u$ in the news network when user $u$ retweets the tweet of user $v$ that contains a URL linking to a website belonging to one of the news media categories. With this convention, the direction of the link represents the direction of the influence between Twitter users. We do not include multiple links with the same direction between the same two users or self-links (when a user retweets their own tweets). The in-degree of a node is the number of links that point inward to the node and its out-degree is the number of links that originate at a node and point outward to other nodes. With our convention, the in-degree of a user is equal to the number of users they retweeted at least once and their out-degree is the number of users who have retweeted them at least once. The higher a node's out-degree, the greater its local influence. The characteristics of the retweet networks are given in SM Tab. S4.

In each retweet network, we use the Collective Influence (CI) algorithm~\cite{Morone2015} to find the best spreaders of information, i.e. the \textit{influencers} of the corresponding news media category. Specifically, the CI algorithm finds the minimal set of nodes that can collectively reach the entire network when information diffuses according to a linear threshold model. The CI algorithm considers influence as an emergent collective property, not as a local property such as the node's out-degree. It does this by finding the smallest set of nodes needed for global cascades. Accordingly, the CI algorithm is able to rank super-spreaders of information in social networks~\cite{Morone2016,Teng2016,Bovet2019}.


\begin{figure*}[htp!]
\centering
\includegraphics[width=0.9\linewidth]{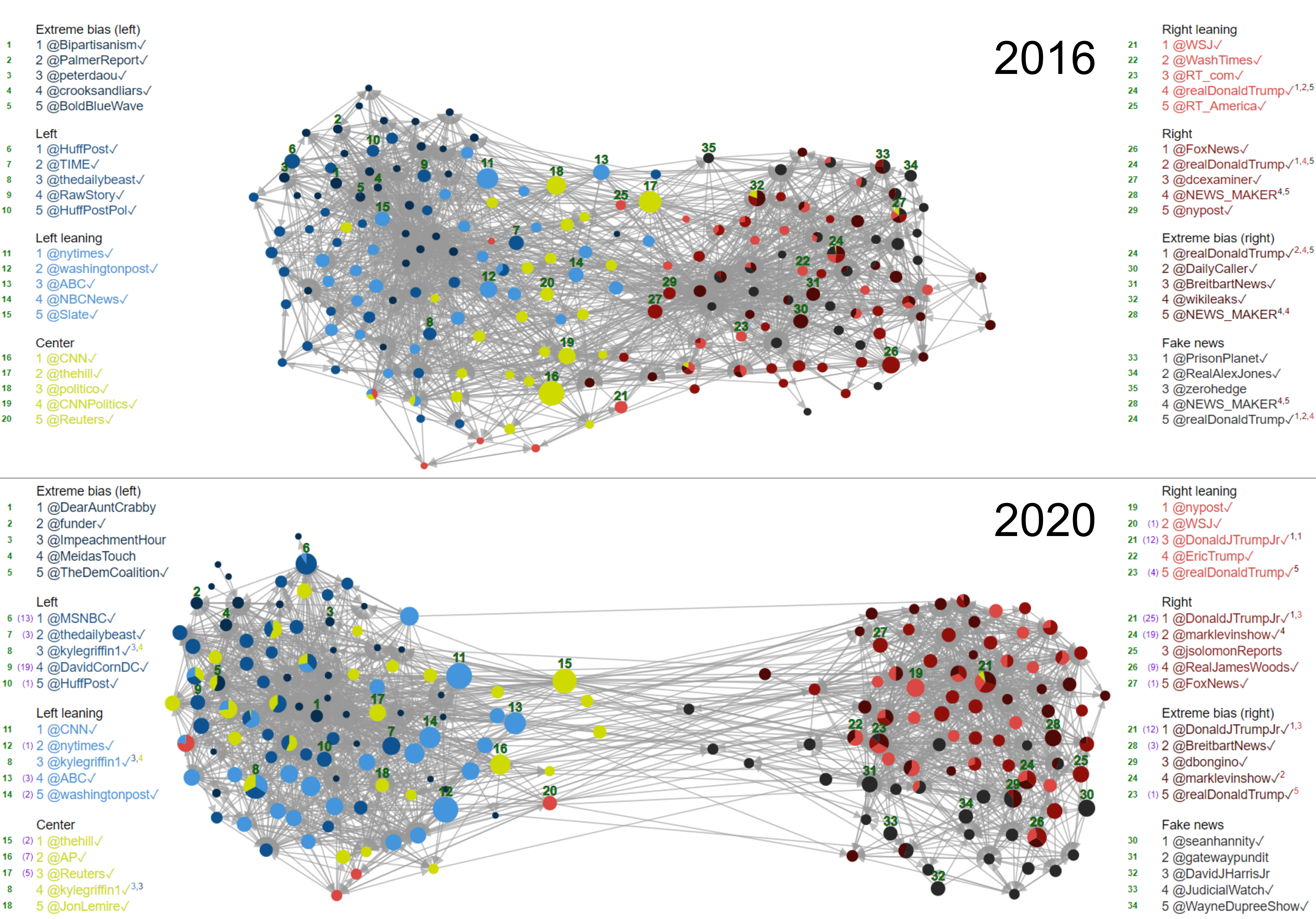}
\caption{\textbf{Retweet networks formed by the top 30 influencers within each media category, by year.} The 2016 network (upper panel) was generated from 2016 data using the same algorithm as used in~\cite{Bovet2019} but with different parameters to ease its comparison with the bottom panel generated from 2020 data. The arrows show the directions of links between users, from the source of influence (the original poster) to the recipient (the retweeter). The size of a node is proportional to its out-degree in the complete combined network, i.e., the number of different users that have retweeted the node at least once with a URL directing to a media outlet. The color of a node indicates the news media category with which the node is affiliated. Nodes ranked in the top 30 of multiple categories are represented by pie charts where the size of each slice is proportional to their CI$_{\textrm{out}}$  ranking (i.e. the node's collective influence). Both networks are visualized using a force-directed graph layout. The tables on either side of the networks show the top five users in each news media category. The number in green to the left of each user is their unique index, used to label the user's node in the network. Users ranked in the top 30 for multiple news media categories have colored superscripts, indicating the rank and media classification of their other top five positions. Verified users are indicated by a checkmark $\checkmark$. In the 2020 tables, a user's 2016 rank is displayed with the purple number to the left of their 2020 rank. Three usernames in the top 10 changed between 2016 and 2020: \texttt{@DRUDGE\_REPORT} became \texttt{@NEWS\_MAKER}, \texttt{@HuffingtonPost} became \texttt{@HuffPost}, and \texttt{@TruthFeedNews} became \texttt{@TAftermath2020}. Those users will have their new handle displayed in the 2016 tables for consistency (as well as in Figs.~\ref{fig:rank_flow_combined} and \ref{fig:latent_ideology}).
}
\label{fig:retweet_networks_combined}
\end{figure*}

Here, we use a directed version of the algorithm to identify the super-spreaders of information as the nodes with the highest CI$_\text{out}$ to be able to compare results from both elections~\cite{Bovet2019}. The 2016 influencers' rankings are shown in the upper panel in Fig.~\ref{fig:retweet_networks_combined} for the top five influencers, and in SM Tab. S5 for the top 25 influencers.  Analysis of these results reveals that traditional news influencers were mostly journalists with verified Twitter accounts linked to traditional news media outlets. In contrast, fake and extremely biased news are sent mainly by influencers whose accounts are unverified or deleted, with deceptive profiles and much shorter life-span on Twitter than traditional media influencers. However, some of these influencers, despite their unknown, non-public nature, still played a major role in the diffusion of disinformation and information on Twitter~\cite{Bovet2019}. 

The results from analysis of 2020 data are shown in the bottom panel of Fig.~\ref{fig:retweet_networks_combined}. For influencers that persisted from 2016, their previous position in 2016 is listed in purple parentheses  (see also SM, Tab. S6). Those influencers are often highly ranked in both the 2016 and the 2020 analyses. Among the union of top 100 influencers from each category in 2020 (representing 598 unique users) 150 were already in the top 100 of one category in 2016. Yet, this means that 75\% of the top 100 influencers in 2020 are new to such high ranking. 

The CI algorithm operates on the unweighted retweet networks. To verify that a ranking computed on the weighted networks would not produce significantly different results, we compare the CI ranking with the ranking obtained from the PageRank algorithm applied to the weighted networks. The comparison reveals a strong agreement, especially for highly-ranked users (SM Fig. S1). 

Fig.~\ref{fig:retweet_networks_combined} shows the retweet networks for each news media category in 2016 and 2020, among communities formed by the union of the top 30 influencers for each category. 
The two force directed network layouts are computed using the same parameters and show two main clusters, with the right-biased and fake news influencers in one cluster and the left-biased influencers in the other. The increased separation in 2020 is notable.
In 2016 the center influencers are mostly between the two clusters; in 2020 the separation between the two clusters increased and only a few influencers remain within a central position (e.g. \texttt{@thehill}). 
We quantify the polarization of the full set of top 100 influencers and of their retweeters, using all the retweets between them, in detail in the next section.

Fake and extremely biased news are sent mostly by influencers whose accounts are unverified or deleted, with fake news seeing a significant increase in deleted influencer accounts from two in the top 25 in 2016 to eight in 2020 (see SM Tabs. S5 and S6). Conversely, the extreme right-biased news in 2020 consisted primarily of verified influencers that grew from 15 verified in the top 25 in 2016 to 23 in 2020.

Using a manual labeling process (see Methods), we classify the top 25 influencers of each news media category for 2016 and 2020 as affiliated with media or political organizations, or unaffiliated, in order to observe the makeup of influencer types for these categories. 
An influencer affiliated with a media organization could be a media company or official media outlet (e.g. \texttt{@FoxNews}), or a writer, reporter, consultant, or other individual who has directly corresponded with a media outlet in an officially recognized capacity (e.g. \texttt{@joelpollak}). An influencer affiliated with a political party could be a politician (e.g. \texttt{@JoeBiden}), a political campaign platform or an affiliate of the platform, or someone who officially represents an aspect of U.S. politics (e.g. \texttt{@joncoopertweets}). We also split the unaffiliated category into two subcategories: independent and ``other.'' An independent influencer is an influencer not officially affiliated with any media or political platforms (e.g. \texttt{@amberofmanyhats}). The ``other'' category represents influencers whose accounts have no descriptions or context that could be used to identify them.

\begin{figure}[htp!]
\centering
\includegraphics[width=0.65\linewidth]{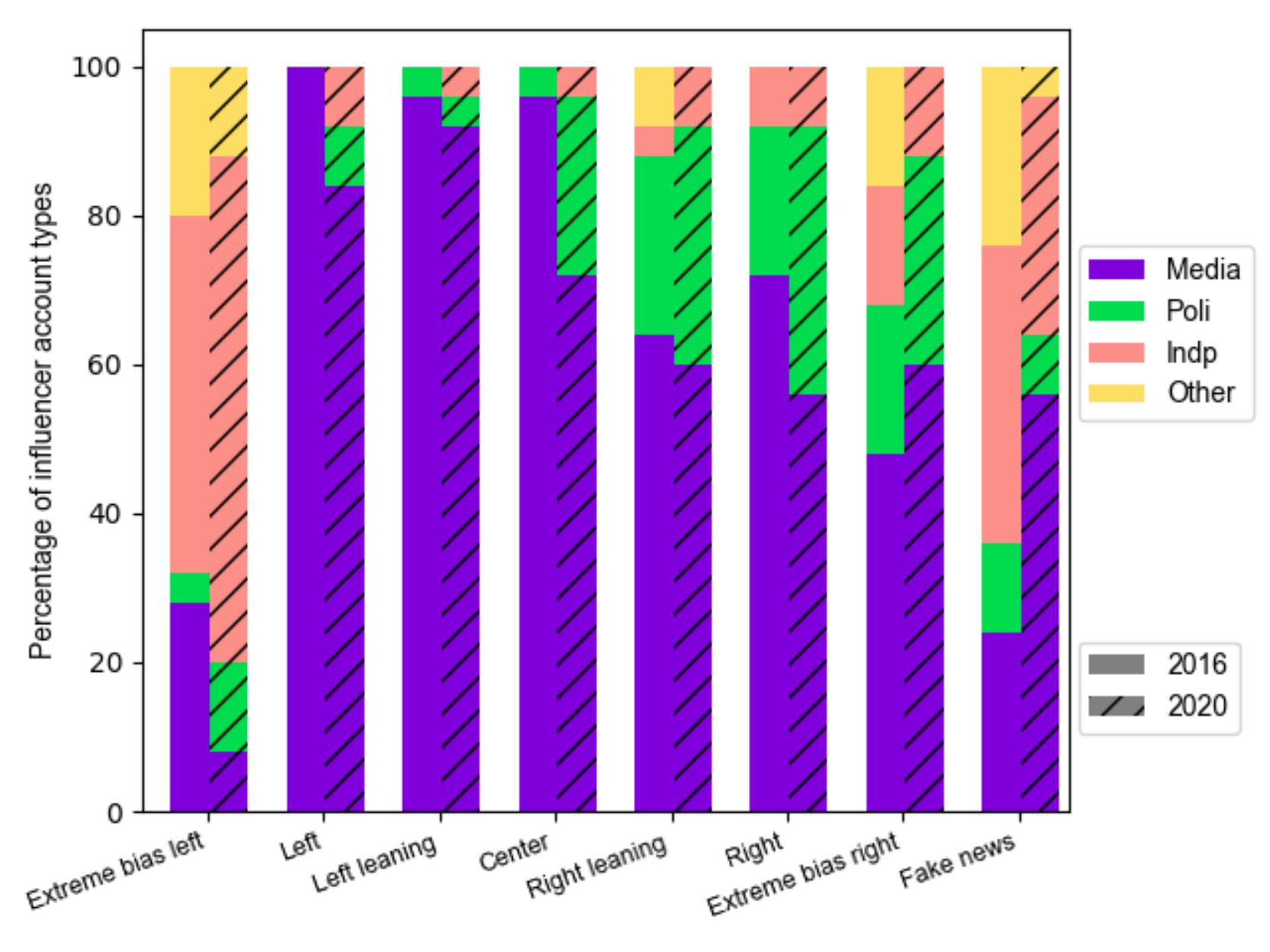}
\caption{
\textbf{Reshuffling  of distribution of the top 25 influencer types from 2016 to 2020, by news media category.} Influencers are classified as affiliated with a media organization, political organization, independent, or other (e.g. unidentified).
}
\label{fig:account_types}
\end{figure}

\begin{figure}[ht]
\centering
\includegraphics[width=\linewidth]{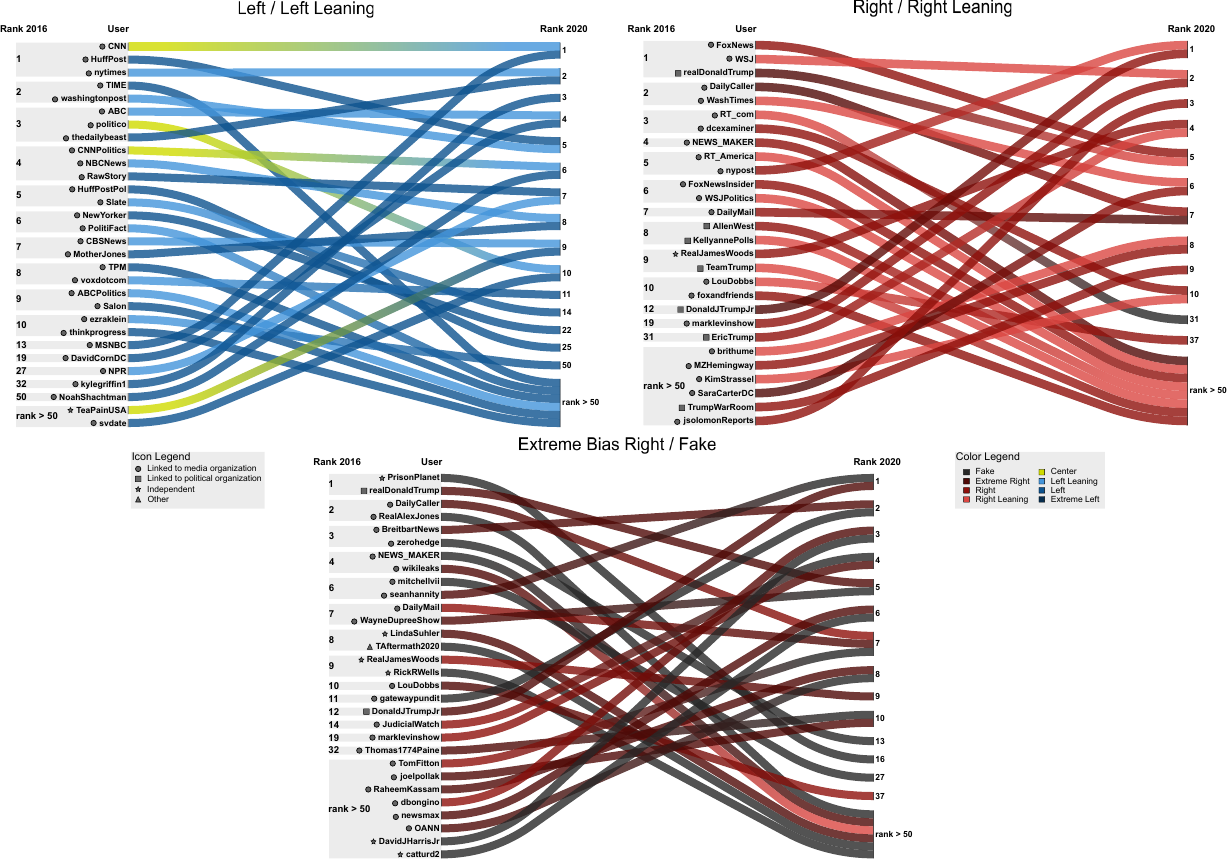}
\caption{
\textbf{Change in influencers rankings from 2016 to 2020.} Influencers ranked in the top 10 in at least one news media category in 2016 or 2020 are shown. The 2016 rankings are displayed to the left of the username, with 2020 rankings on the right. For each user only one shift is shown. Its color changes from the user's highest ranked news media category in 2016 to such color for 2020. Each panel shows the change over time between two news media categories.
\label{fig:rank_flow_combined}
}
\end{figure}

The fractions of influencers in these categories are shown in Fig.~\ref{fig:account_types}. It reveals that unaffiliated influencers are more common in the fake and extreme-bias categories, while affiliated influencers are more common in the other news categories. A similar trend is evident in the fractions of verified and unverified influencers found in these categories (SM Tab. S6), as fake and extreme-bias news categories generally contain fewer verified influencers. In addition, media-affiliated influencers seem to have a greater presence in the left, left-leaning, and center news categories compared to their counterparts. Interestingly, the number of media-affiliated influencers within most of the categories actually decreases from 2016 to 2020. The exceptions are the extreme-right bias and fake news categories, which actually increased in media-affiliated influencers, while the extreme-right bias category also increased in politically-linked influencers. This indicates a shift in polarization of influencers affiliated with right-biased political and media organizations toward the extreme-right bias and fake news, as well as the emergence of new media-affiliated influencers in these categories. We discuss these changes in more detail below. In addition to changes in user types and verified users from 2016 to 2020, we observe a significant reshuffle of the ranking of influencers. Fig.~\ref{fig:rank_flow_combined} shows the change in rankings of the top 10 influencers in left and left-leaning, right and right-leaning and extreme-right bias and fake news categories. The ranking reshuffle in the center news category is shown in SM Fig. S2. 

The comparison reveals several interesting changes between 2016 and 2020. First, we see that highly influential users rise from obscurity. Across all categories, a set of previously unranked or very low-ranked users break in to the top-10 rankings. These users include, for example, \texttt{@TeaPainUSA}, \texttt{@svdate}, \texttt{@kylegriffin1}, \texttt{@marklevinshow}, \texttt{@DavidHarrisJr}, etc. Considering all unique users in the top 25 influential users (across all categories of news media), we see that 58\% came from outside the top 100 influential users in 2016. However, most of these newly influential users are related in some way to media or political organizations, while 28\% of these new influencers are independent. 

Observing the change in rankings by news media category, we see that right/right-leaning and extreme-right bias/fake news categories have a significantly higher fraction of top 10 influencers who were previously outside the top 50, compared to the change in rankings among the groups in left/left-leaning news categories. All categories show a large number of influencers falling out of the top 50 from 2016 to 2020, and in the case of the left news influencers, we see their former positions filled by users who were much less influential in 2016. The influencers with extreme right bias and fake news affiliations show the most volatility with regards to retaining top-10 influencer positions, with many top-10  influencers in 2016 ranked below 50 in 2020 (or were even banned from Twitter, like \texttt{@RealAlexJones}). 

The change of classification of some news media outlets is also reflected in the category shifts of their Twitter accounts.
In particular, \texttt{@CNN} and \texttt{@politico} -- previously the first and third highest ranked influencers in the center category in 2016 -- shifted to left-leaning. 
Such shifts of large and influential media influencers across news categories indicates the increased content polarization on Twitter. 
A shift of media-affiliated influencers from the right to the extreme right is also visible (e.g. \texttt{@DailyMail}, \texttt{@JudicialWatch}, \texttt{@marklevinshow}), as well as the emergence of new-media affiliated influencers in these categories (e.g. \texttt{@newsmax}, \texttt{@OANN} or \texttt{@RaheemKassam}). 
In contrast to the shift to the extremes among large media influencers, the center rankings remained fairly consistent between 2016 and 2020 (SM Fig. S2). Some new users rose from low ranks to fill in the gaps, including \texttt{@JoeBiden}, but only one user dropped out of the top 50 entirely, and the remaining shifts are internal to these top-ranked users.


\subsection*{Polarization among Twitter users}
\label{sec:twitter_pola}

The evolution of influencers across different news media categories (Figs.~\ref{fig:num_tweets_num_users} and \ref{fig:user_flow}) suggests an increased polarization in the relations among influencers between 2016 and 2020. Here we broaden the scope of polarization analysis to the Twitter users who are consuming and retweeting the influencers' content. For the 2016 and 2020 data, we consider the union of the top 100 influencers of each news media category as a single set representing the most influential users covering the entire news media ideological spectrum for the target year. For this analysis we use all the retweets in our datasets, not only those containing a link to a news outlet, and remove the ones sent from unofficial Twitter clients to capture only tweets sent by humans. 
Using these influencers as nodes, we create two fully connected similarity networks derived from the 2016 and 2020 Twitter network, respectively. An edge between any two influencers in the created networks represents similarity of the number of retweets of these two influencers for every user in the corresponding Twitter network (see Methods for more details). In both similarity networks, a community detection algorithm found two communities. One contained influencers affiliated with news media in the center, left-leaning and left news categories, while the other contained those affiliated with news media in the right-leaning, right and fake news categories. This indicates that influencers separate their user bases according to the content they generate. We illustrate this separation in SM Fig. S3 that shows a sample of the 2020 similarity network. To quantify the difference in community separation and, subsequently, polarization, between the two networks, we measured the modularity and normalized cut between communities (see Methods for details). 

The modularity for the 2020 network was 0.39 with a standard error (SE) of 0.01, versus 0.365 (SE = 0.007) in 2016, indicating more closely knit communities in 2020, with stronger in-community ties and weaker between-community ties. 
Consistent with the increase of community modularity, the average normalized cut of 0.36 (SE = 0.04) in 2016 decreased to 0.128 (SE = 0.005) in 2020. To interpret this change, we note that on average, each node in the 2016 similarity network had $64 \%$ of in-community edges and $36 \%$ of across-community edges. The latter fraction decreased to $13 \%$ in 2020, dropping nearly three times lower than it was in 2016. This indicates much stronger separation of these communities in the later election. We also computed the above metrics on networks generated from user quote similarity in order to show that retweets are the strongest form of endorsement of influencer content, and subsequently the best approach for our analysis (see SM Tab. S7).

To quantify and compare the polarization not only among Twitter influencers but also among the users, we infer the ideology of Twitter users based on the ideological alignment of political actors they follow~\cite{Barbera2015a,Barbera2015b}. 
The bipartite network of followers is then projected on a one dimensional scale using correspondence analysis~\cite{benzecri1973analyse,nenadic2007correspondence}, which applies a SVD decomposition of the adjacency matrix standardized to account for the differences in popularity and activity of the influencers and their followers (see Methods for details). Two users are close on the resulting latent ideology scale if they follow similar influencers. This method has been shown to produce ideological estimates of the members of the U.S. Congress highly correlated with ideological estimates based on roll call voting similarity such as DW-NOMINATE~\cite{Barbera2015a}.

For 2016 and 2020, the data for the analysis consists of the union of the top 100 influencers of each news media category and the sets of users that retweeted at least three different influencers (considering all tweets in our datasets, not only the ones with URLs). Following the finding in~\cite{metaxas2015retweets} that ``retweeting indicates not only interest in a message, but also trust in the message and the originator, and agreement with the message contents,'' we interpret  retweeting a form of endorsement of the content being retweeted. 
Twitter offers other types of interactions allowing users to comment on the content, such as quote tweets and replies. 
The ratio of quotes to retweets of users to influencers was very stable and small ($<5\%$) in 2016 and 2020, for users on the left and right sides of the latent ideology (SM Tab. S8A), which motivated our focus on retweets to infer the ideology of users. We note that the ratio of quotes to retweets from users of one side of the ideology spectrum to influencers of the other side increased from 2016 to 2020, indicating an increased usage of quotes to comment on tweets from influencers of the opposite side. However, the overall usage of quotes over retweets remained small (SM Tab. S8B). We extract the coordinates of each user on the first dimension of the results of the correspondence analysis applied to the weighted network of retweets between the users and the influencers (see Methods for the details and robustness checks that we performed). Finally, for 2016 and 2020, the coordinates of all users are standardized to a mean of zero and a standard deviation of one. Two users are close together on the latent ideology scale if they tend to retweet similar influencers. The influencers' latent ideological positions are then computed as the median of their retweeters' positions.

\begin{figure*}[t!]
\centering
\includegraphics[width=0.95\linewidth]{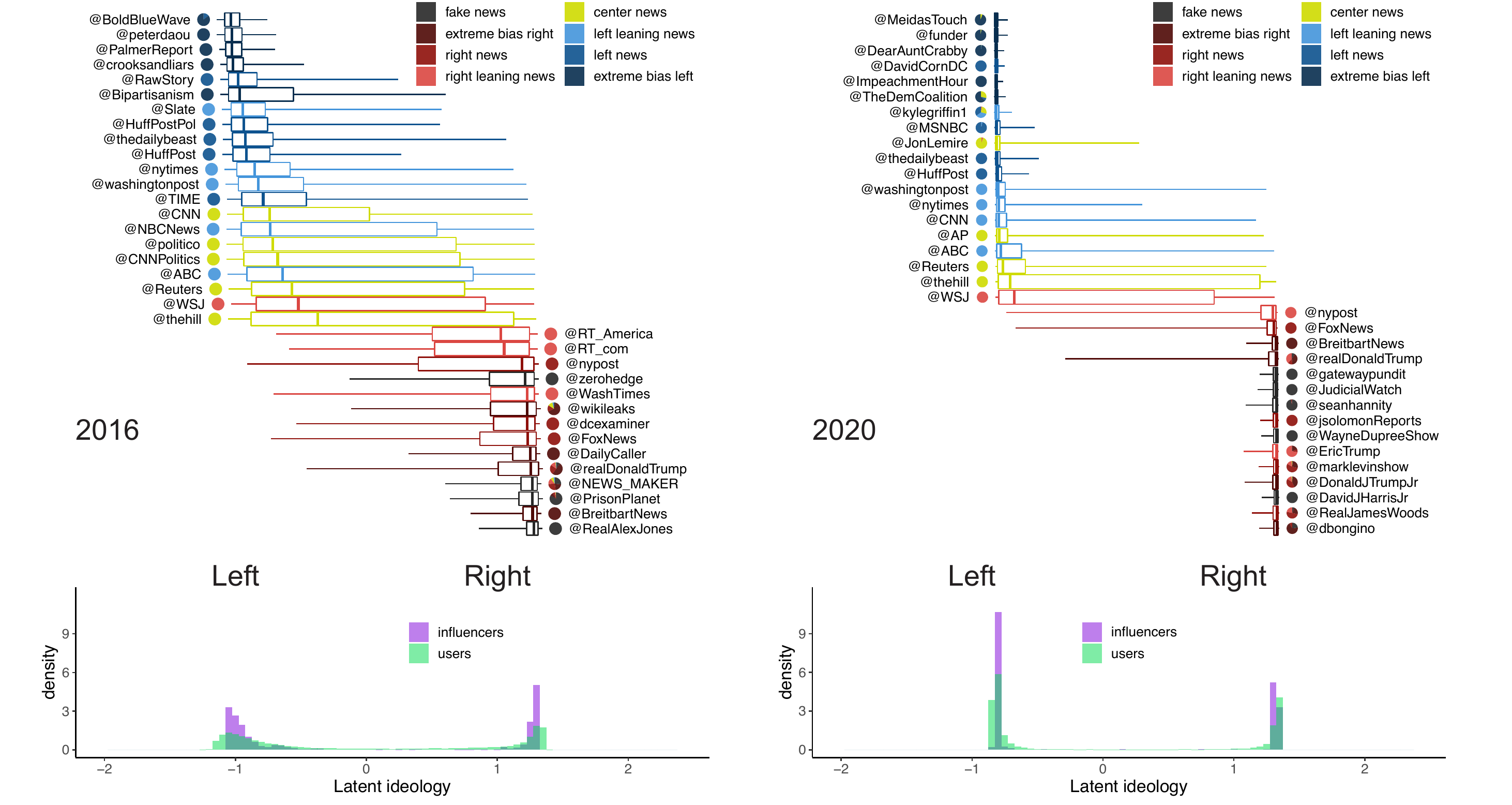}
\caption{\textbf{Latent ideology scale of influencers and their retweeters in 2016 (left) and 2020 (right)}. The latent ideology of the top five influencers of each category is shown as a box plot representing the distribution of the ideology of the users who retweeted them. The distributions for the users are shown in green, and the distributions for the top 100 influencers of each news media category (computed as the median of the ideology of their retweeters) are displayed in purple. Box plots indicate the 25\% and 75\% percentiles of the distributions with whiskers indicating the 5\% and 95\% percentiles. Pie charts next to the influencers' names represent the news categories to which they belong (weighted by their respective CI ranks in each category).}
\label{fig:latent_ideology}
\end{figure*}

Fig.~\ref{fig:latent_ideology} shows the result of this analysis. The latent ideology of the top five influencers of each category is shown as a box plot representing the distribution of the ideology of the users who retweeted them. The distribution of ideology positions of the users and of the influencers, displayed in green and purple, respectively, shows that polarization increased between 2016 and 2020. 
This is confirmed by a Hartigans' dip test (HDT) for unimodality, which measures multimodality in a sample by the maximum difference, over all sample points, between the empirical distribution function, and the unimodal distribution function that minimizes that maximum difference~\cite{hartigan1985dip}.
For the user distribution, the test statistics is $D=0.1086$ (95\% CI: [0.108,0.109]) in 2016 and $D=0.1474$ (95\% CI: [0.1471,0.1477]) in 2020.
For the influencer distribution, the test statistics is $D=0.17$ (95\% CI: [0.16,0.20]) in 2016 and $D=0.21$ (95\% CI: [0.19,0.23]) in 2020. All tests reject the null hypothesis of a unimodal distribution with $p<2.2\times10^{-16}$ and the 95\% confidence intervals are computed from 1000 bootstrap samples using the bias-corrected and accelerated method.
Increasing values of the test statistic indicates distributions that increasingly deviate from a unimodal distribution, corroborating the growing division found in the similarity networks.

To understand if the measured increase in polarization is due to the arrival of new users and influencers in 2020, we repeat this analysis including only users (shown in SM Fig.~S4), only influencers (Fig.~S5) or only users and influencers (Fig.~S6) that were active during both elections.
In all cases we observe an increase of the Hartigans’ dip test (HDT) statistics (see SM Fig.~S7 and Tab.~S9) indicating that the increased polarization is not only due to the departure and arrival of new users between elections but also to a change of behavior of the users that remained.
The largest increase in HDT for the user distribution is obtained when all users of 2016 and 2020 and only influencers that were present during both years are considered ($+0.08$). This setting also corresponds to the smallest increase of the dip test of the influencer distribution ($+0.01$, within 95\% CI), suggesting that the new influencers of 2020 have more polarized ideologies than the influencers who remained from 2016 and that the increased polarization of the users is due in large part to the arrival and departure of users between elections (SM Fig. S7 and SM Tab. S9). 

Figure~\ref{fig:latent_ideology} reveals a clear increase in polarization of the users and influencers in 2020 compared to 2016 and an alignment of their latent ideologies in two distinct groups, mirroring the news media classification groupings seen in Fig.~\ref{fig:retweet_networks_combined} and Fig. S3. The box plots show that the distributions of users retweeting influencers became more concentrated in 2020, with two clear opposite poles and fewer influencers having a user base bridging opposite ideologies. These results independently confirm the shift of news outlets and influencers from the center to the right and left observed using the news media classifications by external sources. Indeed, we find an extremely high correlation (above $0.90$ for 2016 and 2020) between the users' latent ideology position and their left- or right-leaning distribution computed using the news media categories in which they posted (see Methods). This high correlation indicates that the shift in bias observed at the level of the media outlets is also present at the level of the users' retweeting pattern and serves as an independent validation of the media outlet classification.

\section*{Discussion}

This paper uses Twitter retweets to study polarization among influencers and those they influence in the months leading up the 2016 and 2020 U.S. Presidential elections. Multiple analyses confirm a robust pattern of increasing division into opposing echo chambers, largely due to the arrival of new, more polarized influencers and users in 2020. 

Our study also provides unique insights on the rapidly evolving news media landscape on Twitter. Among the top 100 influencers aggregated across all news media categories in 2020, seventy-five percent were not present in 2016, demonstrating how difficult it is to retain influencer status.
Most of the influencers who appeared in 2020 were associated with prominent media or political party organizations. 
The number of influencers affiliated with media organizations declined by 10\% between 2016 and 2020, replaced mostly by influencers affiliated with center and right-leaning political organizations.
This change in the news media landscape on Twitter from 2016 to 2020 indicates a shift in the relative influence of journalists and political organizations.
On the other hand, in the extremely right biased and fake news categories, new professional media have emerged and taken the place of mostly independent influencers.

Future research should build on this structural analysis by examining the content of the messages. Content analysis is needed to distinguish between tweets that are positively and negatively quoted and to develop measures of influence that go beyond the ability to attract attention from retweeters. For example, an urgent question to answer is whether the influence of unaffiliated Twitter influencers goes beyond being news spreaders: do they also have the ability to set the issue agenda? Our study is limited to describing what happened on Twitter. Future research should analyze message content for clues about the ability of influencers to mobilize voters and social movements offline. We also focused on the flow of information from influencers to those who retweet them. Future research should investigate how the actions of the retweeters and followers affect the influencers, how influencers form networks across types of media, and what are the offline consequences of polarization of Twitter influencers and users, including the impact on voting. It should also be possible to monitor interactions on other social media and during non-election periods to permit finer grained analysis of the new entrants. 

\section*{Methods}
\subsection*{News media URL classification}
The website \url{www.allsides.com} (AS) rates media bias using a combination of several methods such as blind surveys, editorial review, third party analysis (e.g. academic research), independent review and community feedbacks (see \url{www.allsides.com/media-bias/media-bias-\\rating-methods} for a detailed explanation of their methodology). 
The website \url{mediabiasfact\\check.com} (MBFC) scores media bias by evaluating wording, sourcing, and story choices as well as political endorsement (see \url{mediabiasfactcheck.com/methodology}).
MBFC is maintained by a small independent team of researchers and journalists, offers the largest set of biased and inaccurate news sources among five fact checking datasets~\cite{Bozarth2020}, and is widely used for labeling bias and veracity of news sources (e.g., in~\cite{main2018rise,stefanov2020predicting,Cinelli2020}).

To be consistent with the results from 2016~\cite{Bovet2019}, we discard as insignificant outlets that accumulate less than $1\%$ of the cumulative number of tweets of the more popular outlets in each category. Removing uniformly insignificant outlets from all categories also ensures that the tweet volume in each category is independent of the number of outlets classified in this category by AS and MBFC. The full lists of outlets in each category in 2016 and 2020 are given in SM Tabs. S1 and S2. AS and MBFC updated their bias classification for several outlets between 2016 and 2020, changing the classification used in our analyses as well. For example, CNN Web News was classified in the \textit{center} category in 2016 by AS and then in the \textit{left-leaning} category in 2020, reflecting a bias shift occurring during this time (see \url{www.allsides.com/blog/yes-cnns-media-\\bias-has-shifted-left}). 


\label{sec:news_classif}
In Ref.~\cite{Bovet2019}, the fake news and extreme-bias categories were based on the classification of a team of media experts (available at \url{github.com/alexbovet/opensources}) and was cross-checked using the factual reporting scores from MBFC. As the classification source from 2016 was not updated in 2020, we use the list of outlets classified as ``questionable sources'' from MBFC as a reference for 2020. MBFC describes a questionable source as one ``that exhibits one or more of the following: extreme bias, consistent promotion of propaganda/conspiracies, poor or no sourcing to credible information, a complete lack of transparency and/or is fake news.'' MBFC rates the factual reporting of each source on a scale from 0 (very high) to 10 (very low) based on their history of reporting factually and backing up claims with well-sourced evidence. Outlets with a level of ``low'' (score of 7, 8 or 9) or ``very low'' (score of 10) are classified in the fake news category while outlets with a ``mixed'' level (score of 5 or 6) are classified in the extremely biased category. No outlets in the disinformation categories have a level higher than ``mixed.''
A ``low'' or ``very low'' factual reporting level on MBCF corresponds to sources that rarely, or almost never use credible sources and ``need to be checked for intentional fake news, conspiracy, and propaganda.'' A ``mixed'' level is assigned to sources that ``do not always use proper sourcing or source to other biased/mixed factual sources.'' We also verify that all outlets in the extremely biased category have a ``bias'' reported on MBFC of ``right'', ``extreme right'', ``left`` or ``extreme left.''

We identify in our datasets (we give the top hostname as an example in parenthesis) for the fake news category: {16} hostnames in 2016 (top: \url{thegatewaypundit.com}) and 20 hostnames in 2020 (top: \url{thegatewaypundit.com}), for the extremely biased (right) category: 17 hostnames in 2016 (top: \url{breitbart.com}) and 10 hostnames in 2020 (top: \url{breitbart.com}), for the extremely biased (left) category: 7 hostnames in 2016 (top: \url{dailynewsbin.com}) and 7 hostnames in 2020 (top: \url{occupydemocrats.com}), for the left news category: 18 hostnames in 2016 (top: \url{huffingtonpost.com}) and 18 hostnames in 2020 (top: \url{rawstory.com}), for the left-leaning news category: 19 hostnames in 2016  (top: \url{nytimes.com}) and 19 hostnames in 2020 (top: \url{nytimes.com}), for the center news category: 13 hostnames in 2016 (top: \url{cnn.com}) and 13 hostnames in 2020 (top: \url{thehill.com}), for the right-leaning news category: 7 hostnames in 2016 (top: \url{wsj.com}) and 13 hostnames in 2020 (top: \url{nypost.com}), for right news category: 20 hostnames in 2016 (top: \url{foxnews.com}) and 
19 hostnames in 2020 (top: \url{foxnews.com}).
The full lists of outlets in each category in 2016 and 2020 are given in SM Tabs. S1 and S2.

\subsection*{Influencer type classification}

For each of the years 2016 and 2020, we manually classified the top-25 influencers in each news media category as affiliated to a media organization or a political organization, or unaffiliated (classified either as an independent user or as an unidentified ``other'' user). The manual labeling procedure was as follows: Eight of the authors were randomly assigned a subset of the union of the top-25 influencers in these category lists to independently classify, such that each subset was examined by three different authors. Each author was shown the account name of the influencer along with descriptions, posts, and all available non-Twitter information such as their Wikipedia entry. Each influencer was then assigned their category based on the majority vote of the three independent classifications.

\subsection*{Similarity network analysis}

We start by creating for each influencer $i$ a vector $\vec{S^i}$ of size $U$, which stands for the number of users in our dataset. We used a set of $588$ influencers for the 2016 dataset, and a set of $661$ influencers for the 2020 dataset. An index $u$ is assigned to the specific user. The vector element $s^i_u$ defines the number of times user $u$ has retweeted influencer $i$. Then, we create the adjacency matrix $\matr{A}$ of size $I\times I$ for our similarity networks by setting $a_{i1,i2}$ to the cosine similarity between vectors $\vec{S^{i_1}}$ and $\vec{S^{i_2}}$. It follows that the higher the cosine similarity, the more users have the similar number of retweets for influencers $i_1,i_2$.

We detect communities in the similarity network using the Louvain algorithm~\cite{blondel2008fast}. In the similarity networks for both election years, we found two communities. Using the accounts of influencers in each community, we found that both election years one community contains influencers primarily associated with fake and right-biased news categories, while the other contains influencers from center and left-biased news categories. This split coincides with an underlying division among the Twitter user bases in the content they propagate.

We quantify the severity of this split using two measures of separation between communities. First is modularity that computes the sum of difference between the fraction of edges within each community and such fraction expected within this community in a random network with the same number of nodes and edges. This metric has a range of $[-0.5,1]$~\cite{brandes2007modularity}. A positive value indicates the presence of communities separated from each other. The closer the modularity is to 1, the stronger communities are separated.  

The second measure uses the normalized cut, which is the sum of the weights of every edge that links a pair communities divided by the sum of the weights of all edges. The result has a range of $[0,1]$ where the smaller the value, the stronger the separation among communities.

\subsection*{Latent ideology estimation}
\label{sec:ca}

The latent ideology estimation follows the method developed in \cite{Barbera2015a,Barbera2015b} adapted to using retweet interactions instead of following relations. As in~\cite{Barbera2015b}, we use correspondence analysis~\cite{benzecri1973analyse} (CA) to infer ideological positions of Twitter users.

The adjacency matrix, $\matr{A}$, of the retweet network between the influencers and their retweeters is the matrix with element $a_{ij}$ equal to the number of times user $i$ retweeted influencer $j$. We only select tweets that have been sent from the official Twitter client in order to limit the presence of bots and professional accounts and we also remove users that show a low interest in the U.S. elections by removing users that retweeted less than three different influencers. For the 2016 data, the matrix $\matr{A}$ has 751,311 rows corresponding to distinct users, 593 columns corresponding to influencers and the total number of retweets equal to 39,385,772. For the 2020 data, the matrix $\matr{A}$ has 2,034,970 rows corresponding to distinct users, 591 columns corresponding to influencers and the total number of retweets equal to 153,463,788.

The CA method is executed in the following steps~\cite{nenadic2007correspondence}. The matrix of standardized residuals of the adjacency matrix is computed as $\matr{S}=\matr{D}^{-1/2}_r(\matr{P}-\matr{r}\matr{c})\matr{D}^{-1/2}_c$, where $\matr{P}=\matr{A}(\sum_{ij}a_{ij})^{-1}$ is the adjacency matrix normalized by the total number of retweets, $\matr{r}=\matr{P}\matr{1}$ is the vector of row sums, $\matr{c}=\matr{1}^T\matr{P}$ is the vector of column sums, $\matr{D}_r=\text{diag}(\matr{r})$ and $\matr{D}_c=\text{diag}(\matr{c})$. Using the standardized residuals allows the inference to account for the variation of popularity and activity of the influencers and the users, respectively~\cite{Barbera2015b}. Then, a SVD is computed such that $\matr{S}=\matr{U}\matr{D}_{\alpha}\matr{V}^T$ with $\matr{U}\matr{U}^T=\matr{V}\matr{V}^T=\matr{I}$ and $\matr{D}_\alpha$ being a diagonal matrix with the singular values on its diagonal. The positions of the users are given by the standard row coordinates: $\matr{X}=\matr{D}_r^{-1/2}\matr{U}$ where we only consider the first dimension, corresponding to the largest singular value. Finally, the ideological positions of the users are found by standardizing the row coordinates to have a mean of zero and a standard deviation of one. The ideological position of the influencers is given by the median of the weighted positions of their retweeters.

We tested the robustness of our method by varying the way we construct matrix $\matr{A}$ as follow: 1) removing entries with weight 1 to discard relations showing a weak ideological alignment; 2) considering the logarithm of the number of retweets as weight for influencer for a sublinear relation between the number of retweets and the strength of ideology alignment; 3) considering a random subsample of the 2020 retweet data of the same size than the 2016 retweet data to control for a potential effect of the difference in sizes of the two datasets. All of these robustness tests match the results of our initial method with correlation coefficients between the user position distributions in the robustness tests and in the initial configuration at above 0.995. We also compare the users' latent ideology distribution with the users average leaning distribution and find a correlation above 0.90 for 2016 and 2020. The average leaning of users is computed for all users having at least three tweets classified in at least one news media category and estimated as the weighted average of the news media category positions, given as: fake news = 4/3, extreme-right bias = 1, right = 2/3, right-leaning = 1/3, center = 0, left-leaning = -1/3, left = -2/3, extreme-left bias = -1.

\bibliography{ms}
\bibliographystyle{ieeetr}

\subsection*{Author Contributions}
SF, HAM and BKS conceived the research; BKS, HAM and AB designed and supervised the research; AB and JF coordinated and supervised the analysis; SF, MWM, AB, HAM and BKS defined the scope of the paper; JF, AG, BC, ZZ, MS and AB processed the collected data; all authors analyzed the results' AB, JF, MWM, and BKS wrote the first draft; and all authors edited and approved the paper.

\subsection*{Author Declaration}
All authors declare no competing interests.

\subsection*{Acknowledgments}
JF, BC, and BKS were partially supported by the Army Research Office (ARO) under Grant W911NF-16-1-0524 and by DARPA under Agreements W911NF-17-C-0099 and HR001121C0165. HAM was supported by NSF DMR-1945909.

\end{document}


\baselineskip24pt

\maketitle 


\begin{figure}[hp!]
    \centering
    \includegraphics[width=0.95\textwidth]{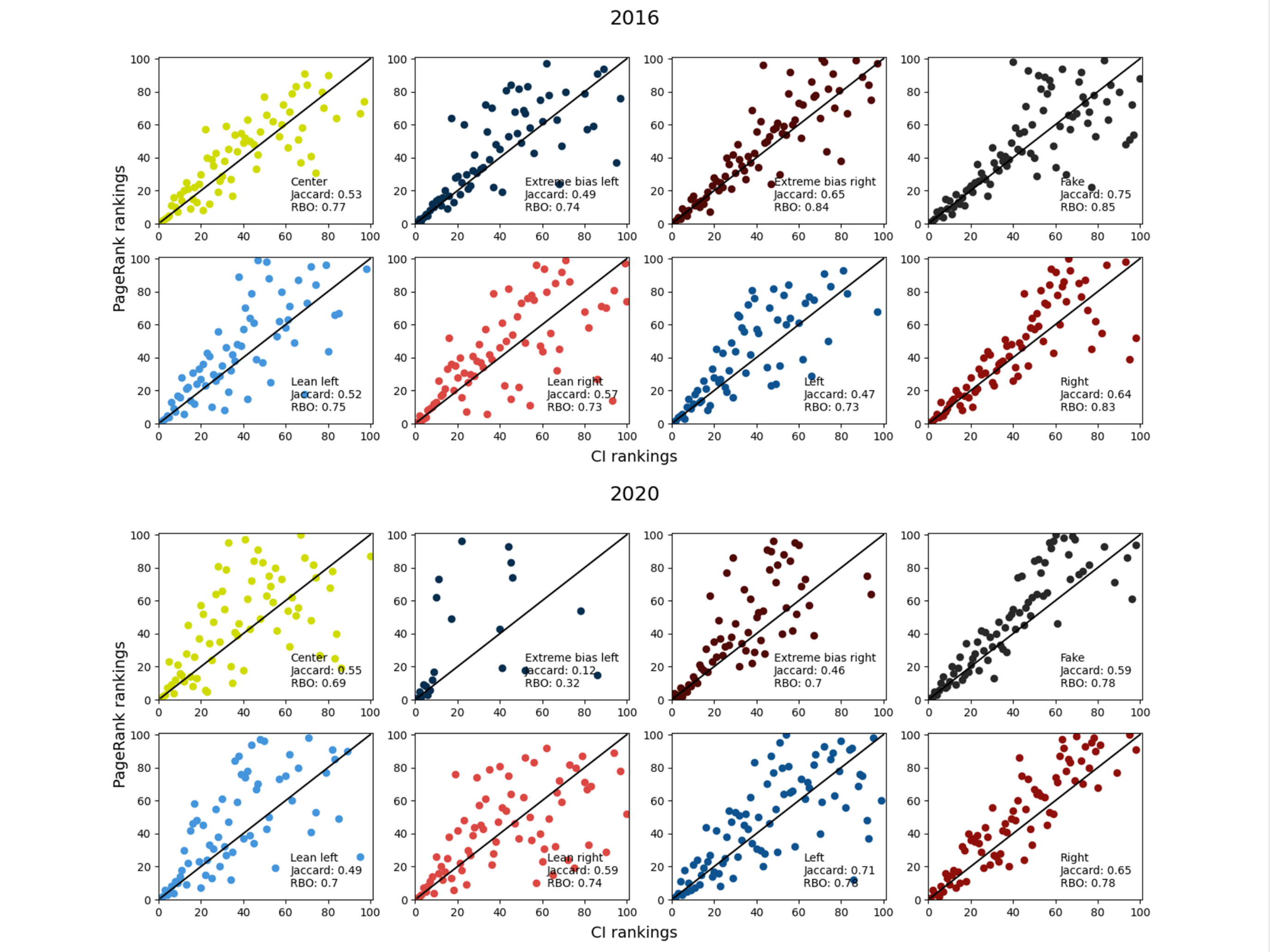}
    \caption{
    \textbf{Comparison of top 100 rankings generated by the PageRank algorithm and by the Collective Influence (CI) algorithm using the 2016 and 2020 retweet networks.} Ranked Bias Overlap (RBO) \cite{webber2010similarity} and Jaccard Similarity are computed over the two top 100 lists, shown below their respective news category labels. For this analysis, RBO's weight parameter $p$ is set to $0.98$. The RBO values are generally above 0.7 indicating a high agreement of the two ranking, especially for the top ranked users. The only network that show a poor agreement between the rankings is the extreme bias left network of 2020. This may be explained by the small size and low average degree of the network compared to networks of other categories (see Tab. \ref{tab:retweet_net}).}
    \label{fig:ci_vs_pr}
\end{figure}

\newpage

\begin{figure}[hp!]
    \centering
    \includegraphics[width=0.5\textwidth]{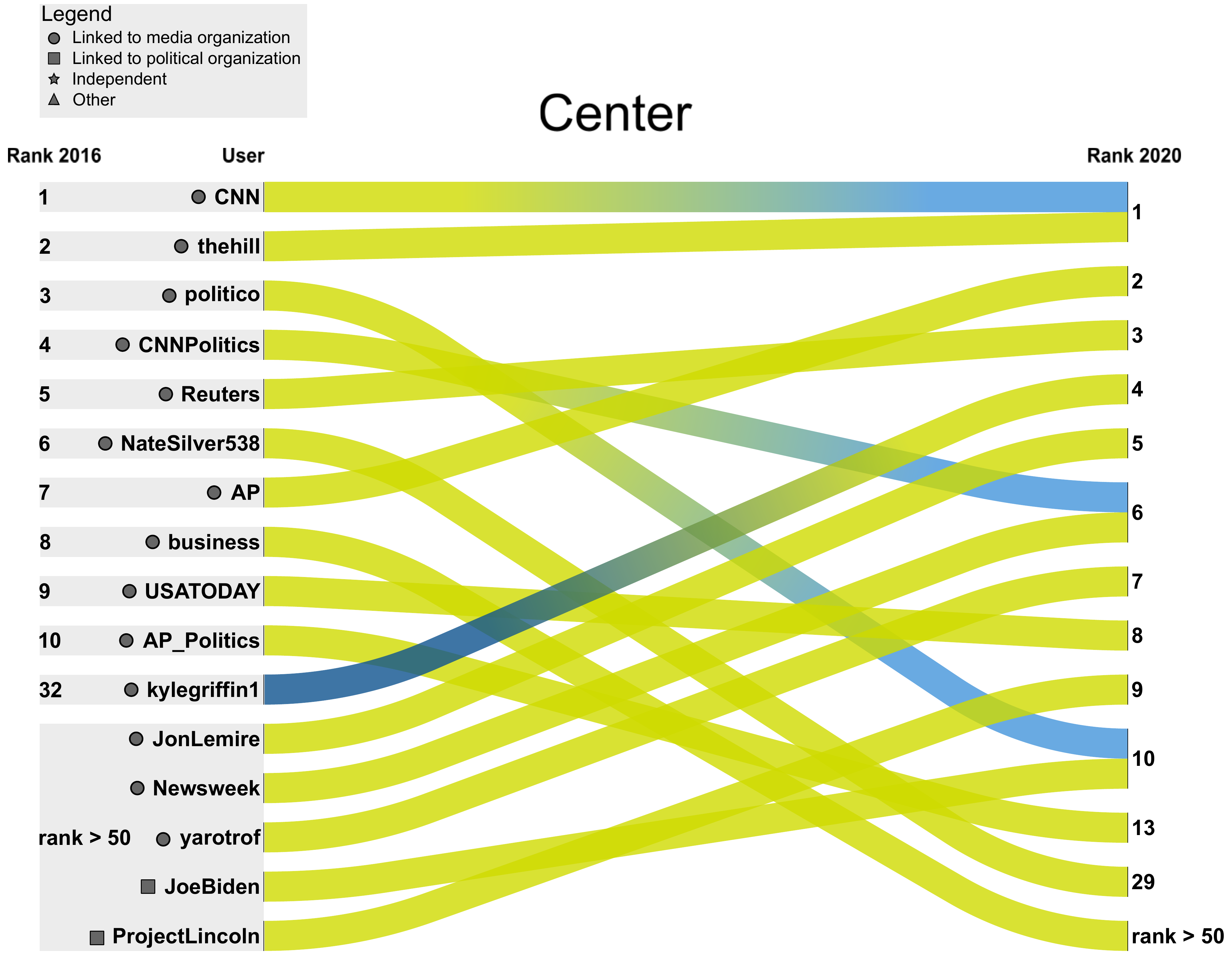}
    \caption{\textbf{Change in rankings 2016-2020, Center Bias}.
Outlines the change in the ranks of the top 10 center bias users from 2016 and 2020, ranked by CI influence. Each flow connects the best ranking for a user in 2016, whose rank is displayed to the left of the user handle, to their rank in 2020. The color of the lines match the bias of the users best ranking, and gradients represent a change in the bias classification of their best ranking. Note user @kylegriffin1 is more highly ranked in the left leaning bias (rank 3) but we chose to show its center ranking for this center bias plot, as the difference in rank is small and it keeps the figure focused on the center bias.}
    \label{fig:change_in_rankings_center}
\end{figure}

\newpage

\begin{figure}
\centering
\includegraphics[width=0.9\linewidth]{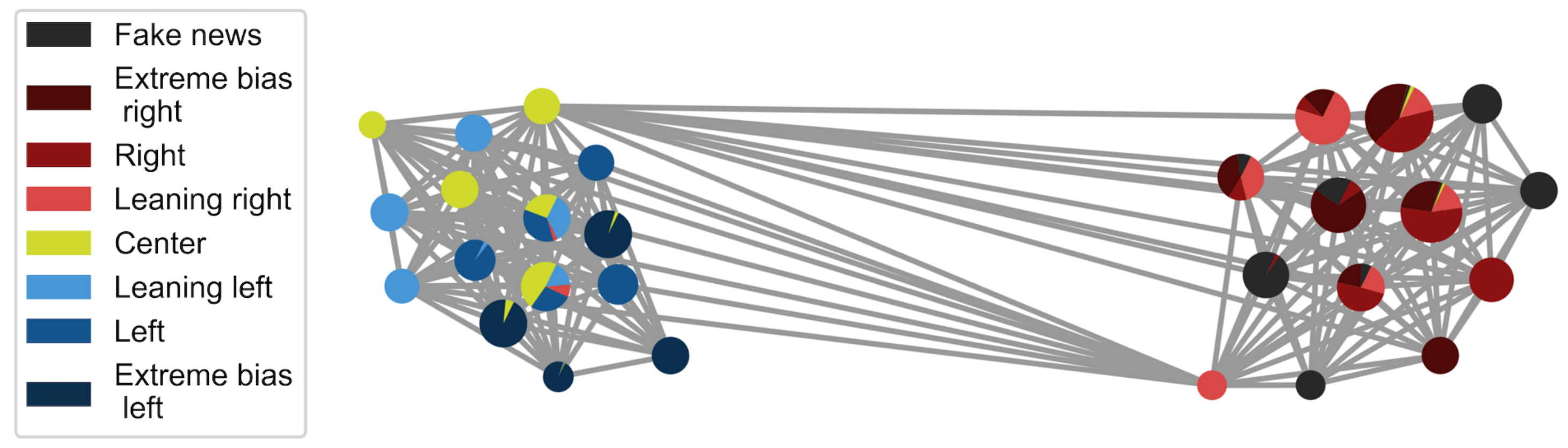}
\caption{\textbf{Similarity network for a random subsample of the 2020 influencers}. Each edge is weighted by the cosine similarity between retweeting users. Size of the node represents that node's degree centrality. The pie charts representing the nodes illustrate the news categories to which that node belongs, with the size of the slices denoting their relative influence for that category. For clarity, edges below the average inter-community edge weight are hidden. Nodes are grouped relative to each other by their detected community.
}
\label{fig:community_network}
\end{figure}

\begin{figure}
    \centering
    \includegraphics[width=0.95\textwidth]{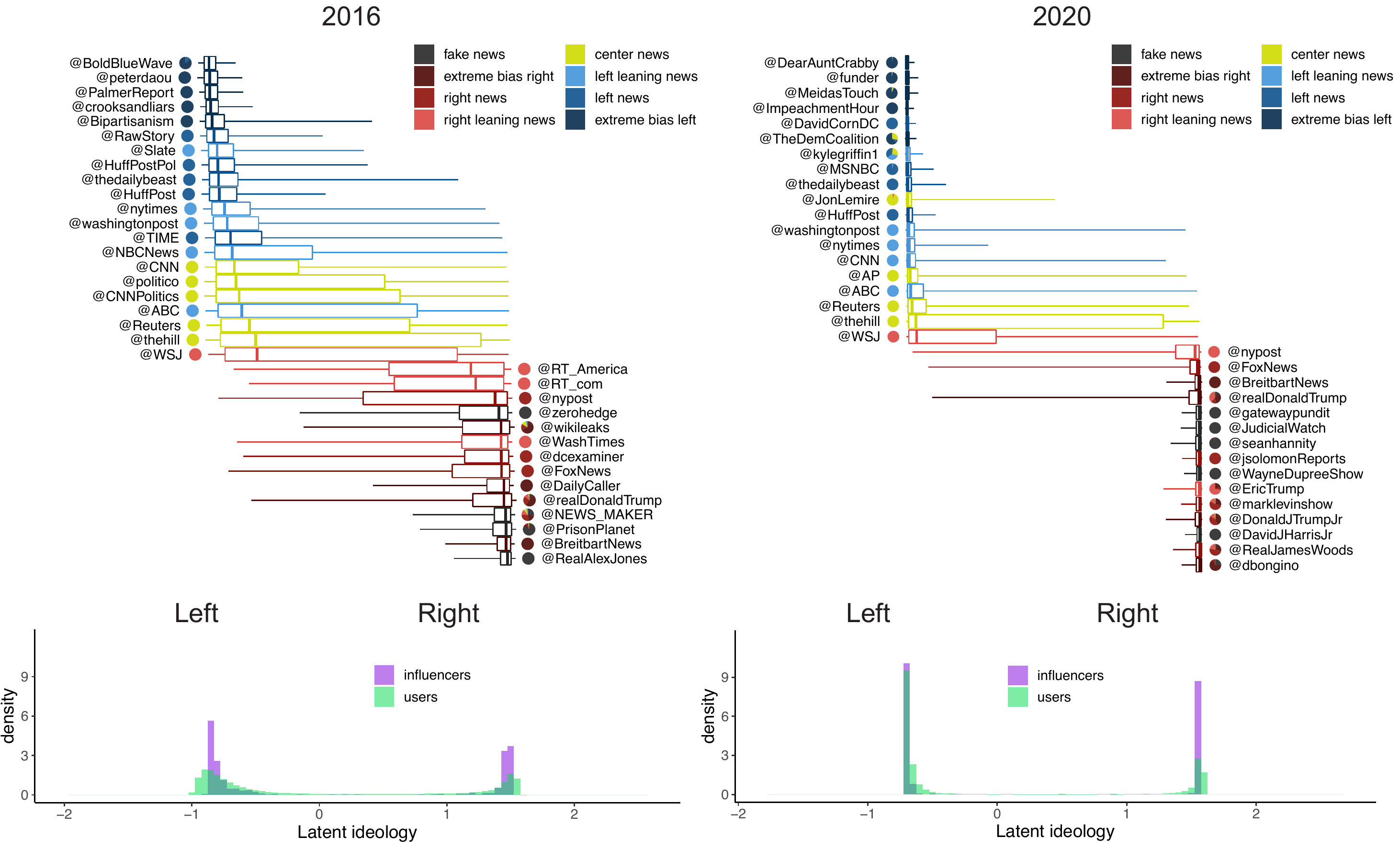}
    \caption{\textbf{Latent ideology scale of influencers and their retweeters in 2016 (left) and 2020 (right) using only users active in both years}.
The latent ideology of the top 5 influencers of each category is shown as a box plot representing the distribution of the ideology of the users having retweeted them.
The distribution of the ideology estimate of the users is shown in green and the 
distribution of the ideology estimate of the top 100 influencers of each news category (computed as the median of the ideology of their retweeters) is displayed in purple.
Pie charts next to the influencers' names represent the news categories they belong to (weighted by their respective CI ranks in each category). Hartigans’ dip test for unimodality applied to the user distribution is $D =  0.094$ ($p< 2.2\times10^{-16}$) in 2016 and $D = 0.117$ ($p< 2.2\times10^{-16}$) in 2020.
The test statistics for the influencer distribution is $D = 0.178$ ($p< 2.2\times10^{-16}$) in 2016 and $D = 0.214$ ($p< 2.2\times10^{-16}$) in 2020.
}
    \label{fig:common_users_ideology}
\end{figure}

\begin{figure}
    \centering
    \includegraphics[width=0.95\textwidth]{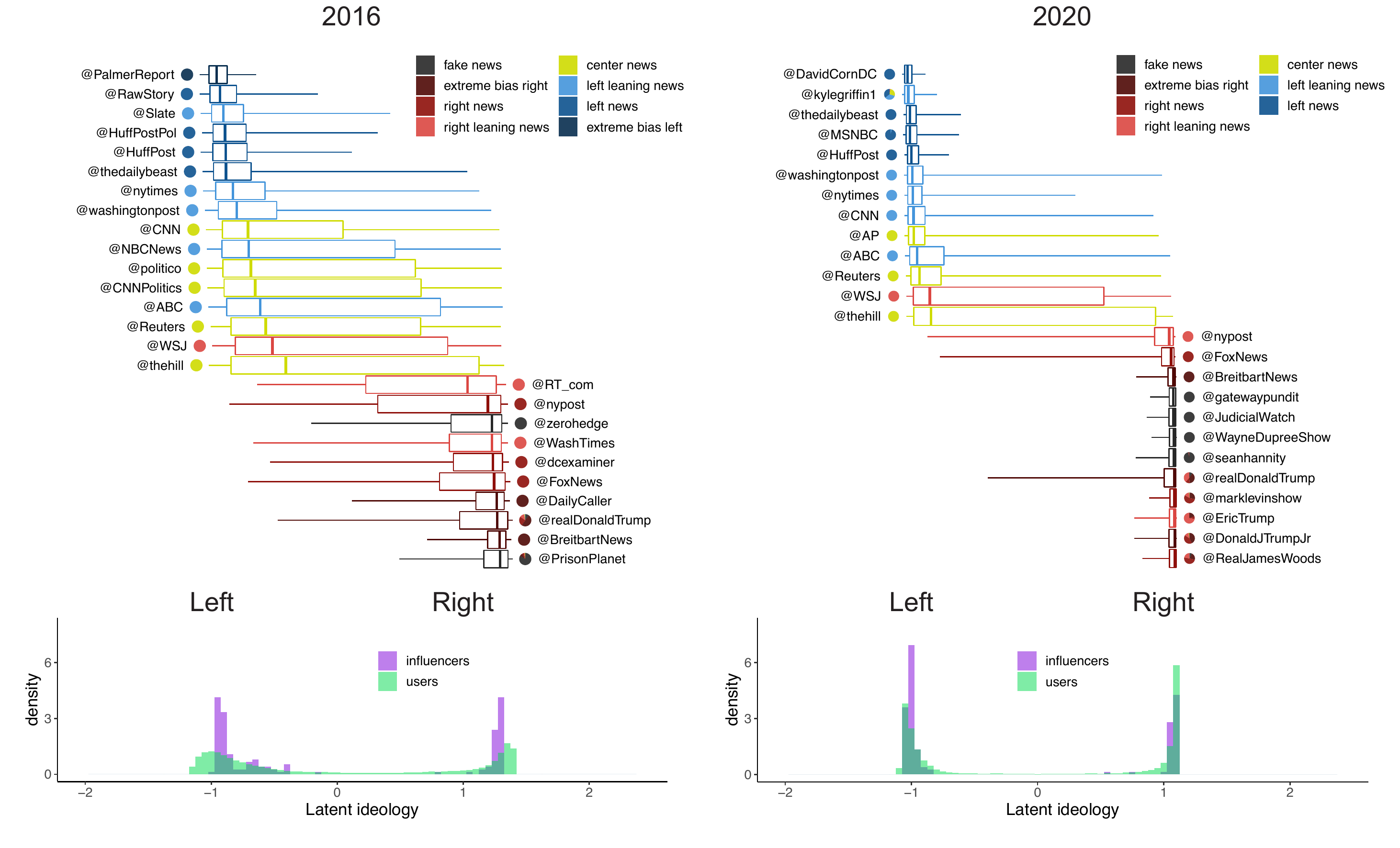}
    \caption{\textbf{Latent ideology scale of influencers and their retweeters in 2016 (left) and 2020 (right) using only influencers active in both years}.
The latent ideology of the top 5 influencers of each category is shown as a box plot representing the distribution of the ideology of the users having retweeted them.
The distribution of the ideology estimate of the users is shown in green and the 
distribution of the ideology estimate of the top 100 influencers of each news category (computed as the median of the ideology of their retweeters) is displayed in purple.
Pie charts next to the influencers' names represent the news categories they belong to (weighted by their respective CI ranks in each category).
Hartigans’ dip test for unimodality applied to the user distribution is $D = 0.107$ ($p< 2.2\times10^{-16}$) in 2016 and $D = 0.183$ ($p< 2.2\times10^{-16}$) in 2020.
The test statistics for the influencer distribution is $D = 0.163$ ($p< 2.2\times10^{-16}$) in 2016 and $D = 0.173$ ($p< 2.2\times10^{-16}$) in 2020.}
    \label{fig:common_influencers_ideology}
\end{figure}

\begin{figure}
    \centering
    \includegraphics[width=0.95\textwidth]{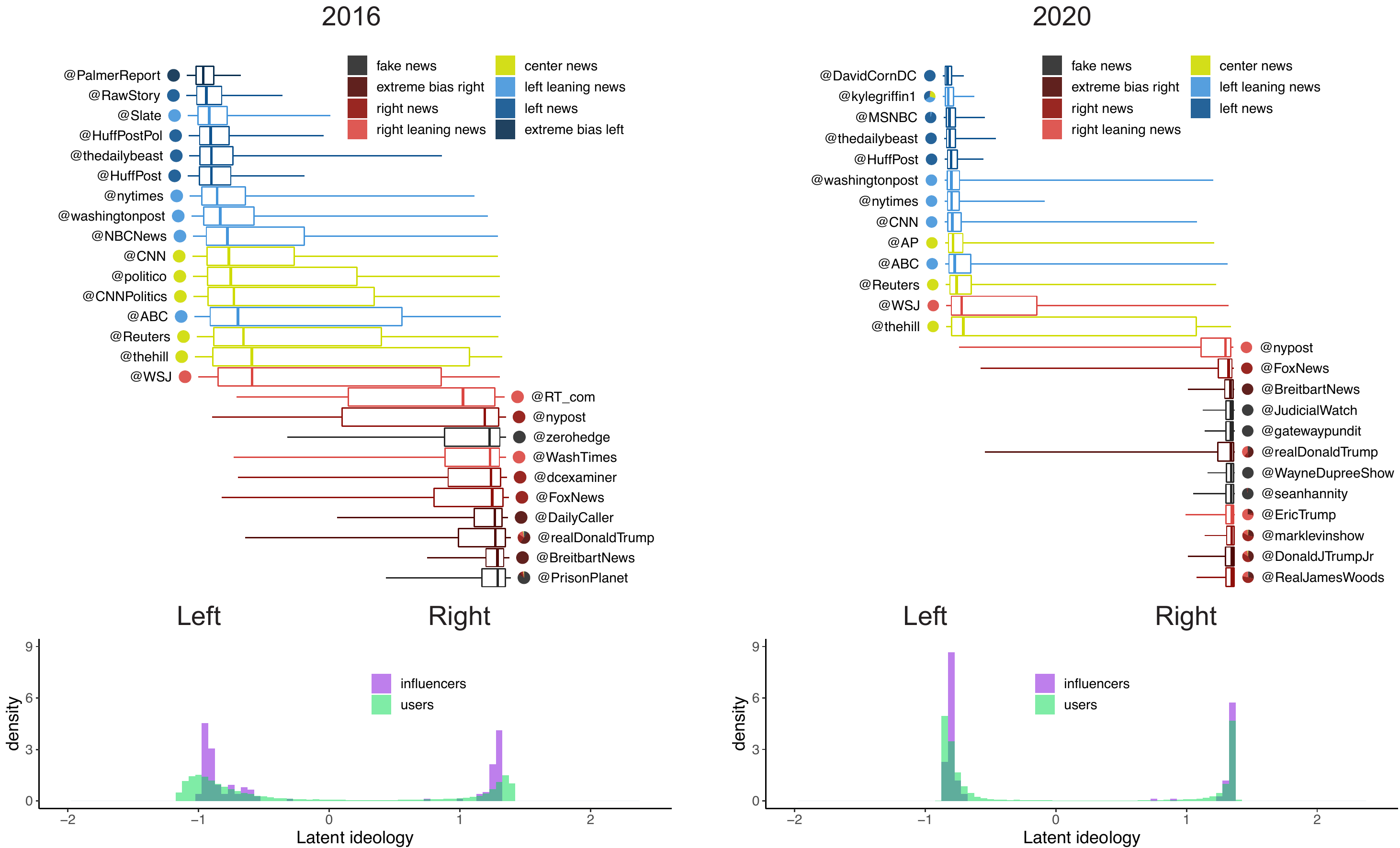}
    \caption{\textbf{Latent ideology scale of influencers and their retweeters in 2016 (left) and 2020 (right) using only users and influencers active in both years}.
The latent ideology of the top 5 influencers of each category is shown as a box plot representing the distribution of the ideology of the users having retweeted them.
The distribution of the ideology estimate of the users is shown in green and the 
distribution of the ideology estimate of the top 100 influencers of each news category (computed as the median of the ideology of their retweeters) is displayed in purple.
Pie charts next to the influencers' names represent the news categories they belong to (weighted by their respective CI ranks in each category).
Hartigans’ dip test for unimodality applied to the user distribution is $D = 0.095$ ($p< 2.2\times10^{-16}$) in 2016 and $D = 0.140$ ($p< 2.2\times10^{-16}$) in 2020.
The test statistics for the influencer distribution is $D = 0.164$ ($p< 2.2\times10^{-16}$) in 2016 and $D = 0.171$ ($p< 2.2\times10^{-16}$) in 2020.}
    \label{fig:common_users_influencers_ideology}
\end{figure}

\begin{figure}
\centering
\includegraphics[width=0.75\textwidth]{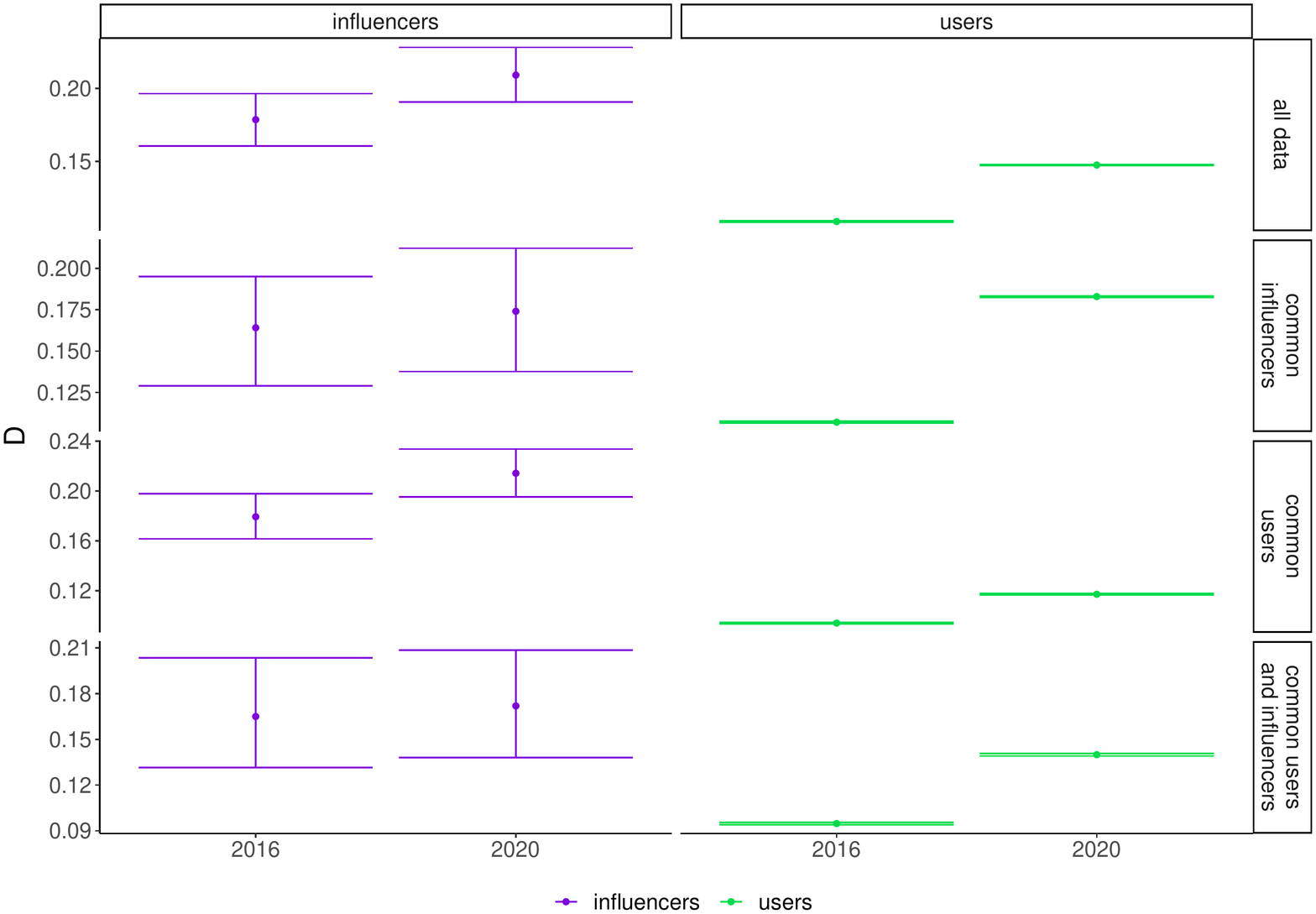}
\caption{Hartigans' dip test values for ideology distribution of users and influencers when considering all users and influencers or only influencers or users present in 2016 and 2020. 95\% CI error bars are obtained by bootstrap with 1000 runs for each dataset and Bias-corrected and accelerated confidence intervals method. The numerical values are reported in Table \ref{tab:dip_tests}.}
\label{fig:dip_tests}
\end{figure}

\begin{table*}[t!]
\centering
\begin{adjustbox}{width=0.65\columnwidth,center}

 \begin{tabular}{rlS[table-format = 6]
                  lS[table-format = 6]
                  lS[table-format = 6]}
  {} &
      \multicolumn{2}{c}{fake news} &
      \multicolumn{2}{c}{extreme bias (right) news} &
      \multicolumn{2}{c}{right news} \\
  {} &            {hostnames} &  {$N$} &                       {hostnames} &  {$N$} &                   {hostnames} &  {$N$} \\
\midrule
1  &     thegatewaypundit.com &  761756 &          breitbart.com &      1854920 &                   foxnews.com &  1122732 \\
2  &            truthfeed.com &  554955 &        dailycaller.com &       759504 &               dailymail.co.uk &   474846 \\
3  &             infowars.com &  478872 &    americanthinker.com &       179696 &        washingtonexaminer.com &   462769 \\
4  &      therealstrategy.com &  241354 &                wnd.com &       141336 &                    nypost.com &   441648 \\
5  &  conservativetribune.com &  212273 &         freebeacon.com &       129077 &              bizpacreview.com &   170770 \\
6  &            zerohedge.com &  186706 &      newsninja2012.com &       127251 &            nationalreview.com &   164036 \\
7  &             rickwells.us &   78736 &            hannity.com &       114221 &                 lifezette.com &   139257 \\
8  &              departed.co &   72773 &            newsmax.com &        94882 &                  redstate.com &   105912 \\
9  &  thepoliticalinsider.com &   66426 &       endingthefed.com &        88376 &                allenbwest.com &   104857 \\
10 &        therightscoop.com &   63852 &         truepundit.com &        84967 &  theconservativetreehouse.com &   102515 \\
11 &             teaparty.org &   48757 &  westernjournalism.com &        77717 &                  townhall.com &   102408 \\
12 &       usapoliticsnow.com &   46252 &          dailywire.com &        67893 &                 investors.com &   102295 \\
13 &           clashdaily.com &   45970 &        newsbusters.org &        60147 &                  theblaze.com &    99029 \\
14 &  thefederalistpapers.org &   45831 &     ilovemyfreedom.org &        54772 &         theamericanmirror.com &    91538 \\
15 &          redflagnews.com &   45423 &    100percentfedup.com &        54596 &                       ijr.com &    71558 \\
16 &     thetruthdivision.com &   44486 &            pjmedia.com &        46542 &             judicialwatch.org &    70543 \\
17 &                      { } &     { } &       weaselzippers.us &        45199 &             thefederalist.com &    55835 \\
18 &                      { } &     { } &                    { } &          { } &                    hotair.com &    55431 \\
19 &                      { } &     { } &                    { } &          { } &        conservativereview.com &    54307 \\
20 &                      { } &     { } &                    { } &          { } &            weeklystandard.com &    50707 \\
\bottomrule
\end{tabular}
\end{adjustbox}
\vspace{0.1cm}

\begin{adjustbox}{width=0.65\columnwidth,center}
 \begin{tabular}{rlS[table-format = 6]
                  lS[table-format = 6]
                  lS[table-format = 6]}
  {} &
      \multicolumn{2}{c}{right leaning news} &
      \multicolumn{2}{c}{center news} &
      \multicolumn{2}{c}{left leaning news} \\
  {} &            {hostnames} &  {$N$} &                       {hostnames} &  {$N$} &                   {hostnames} &  {$N$} \\
\midrule
1  &                wsj.com &        310416 &              cnn.com &   2291736 &            nytimes.com &      1811627 \\
2  &    washingtontimes.com &        208061 &          thehill.com &   1200123 &     washingtonpost.com &      1640088 \\
3  &                 rt.com &        157474 &         politico.com &   1173717 &            nbcnews.com &       512056 \\
4  &  realclearpolitics.com &        128417 &         usatoday.com &    326198 &         abcnews.go.com &       467533 \\
5  &        telegraph.co.uk &         82118 &          reuters.com &    283962 &        theguardian.com &       439580 \\
6  &             forbes.com &         64186 &        bloomberg.com &    266662 &                vox.com &       369789 \\
7  &            fortune.com &         57644 &  businessinsider.com &    239423 &              slate.com &       279438 \\
8  &                    { } &           { } &           apnews.com &    198140 &           buzzfeed.com &       278642 \\
9  &                    { } &           { } &         observer.com &    128043 &            cbsnews.com &       232889 \\
10 &                    { } &           { } &  fivethirtyeight.com &    124268 &         politifact.com &       198095 \\
11 &                    { } &           { } &              bbc.com &    118176 &            latimes.com &       190994 \\
12 &                    { } &           { } &          ibtimes.com &     72424 &        nydailynews.com &       188769 \\
13 &                    { } &           { } &            bbc.co.uk &     71941 &        theatlantic.com &       177637 \\
14 &                    { } &           { } &                  { } &       { } &           mediaite.com &       152877 \\
15 &                    { } &           { } &                  { } &       { } &           newsweek.com &       149490 \\
16 &                    { } &           { } &                  { } &       { } &                npr.org &       142143 \\
17 &                    { } &           { } &                  { } &       { } &      independent.co.uk &       127689 \\
18 &                    { } &           { } &                  { } &       { } &                 cnb.cx &        87094 \\
19 &                    { } &           { } &                  { } &       { } &  hollywoodreporter.com &        84997 \\
\bottomrule
\end{tabular}
\end{adjustbox}
\begin{adjustbox}{width=0.5\columnwidth,center}
\vspace{0.1cm}

 \begin{tabular}{rlS[table-format = 6]
                  lS[table-format = 6]}
  {} &
      \multicolumn{2}{c}{left news} &
      \multicolumn{2}{c}{extreme bias (left) news} \\
  {} &            {hostnames} &  {$N$} &                       {hostnames} &  {$N$} \\
\midrule
1  &     huffingtonpost.com & 1057518 &      dailynewsbin.com &      189257 \\
2  &      thedailybeast.com &  378931 &  bipartisanreport.com &      119857 \\
3  &           dailykos.com &  324351 &  bluenationreview.com &       75455 \\
4  &           rawstory.com &  297256 &    crooksandliars.com &       73615 \\
5  &       politicususa.com &  293419 &   occupydemocrats.com &       73143 \\
6  &               time.com &  252468 &         shareblue.com &       50880 \\
7  &        motherjones.com &  210280 &           usuncut.com &       27653 \\
8  &  talkingpointsmemo.com &  199346 &                   { } &         { } \\
9  &              msnbc.com &  177090 &                   { } &         { } \\
10 &           mashable.com &  173129 &                   { } &         { } \\
11 &              salon.com &  172807 &                   { } &         { } \\
12 &      thinkprogress.org &  172144 &                   { } &         { } \\
13 &          newyorker.com &  171102 &                   { } &         { } \\
14 &       mediamatters.org &  152160 &                   { } &         { } \\
15 &              nymag.com &  121636 &                   { } &         { } \\
16 &       theintercept.com &  109591 &                   { } &         { } \\
17 &          thenation.com &   54661 &                   { } &         { } \\
18 &             people.com &   47942 &                   { } &         { } \\
\bottomrule

\end{tabular}
\end{adjustbox}
\caption{\textbf{Hostnames in each media category in 2016}.
We also show the number ($N$) of tweets 
with a URL pointing toward each hostname. 
Tweets with several URLs are counted multiple times.
Reproduced from \cite{Bovet2019}.}
\label{tab:hostnames2016}
\end{table*}

\begin{table*}
\centering
\begin{adjustbox}{width=0.65\columnwidth,center}

\begin{tabular}{rllS[table-format = 6]llS[table-format = 6]llS[table-format = 6]}
\multicolumn{1}{l}{} &  & \multicolumn{2}{c}{fake news}         &  & \multicolumn{2}{c}{extreme bias (right) news} &  & \multicolumn{2}{c}{right news}   \\
                     &  & hostnames                   & $N$      &  & hostnames                   & $N$              &  & hostnames             & $N$       \\ \toprule
 1 & & thegatewaypundit.com &  1883852 & & breitbart.com & 2192997 & & foxnews.com & 3136578 \\ 
 2 & & hannity.com & 428483 & & dailymail.co.uk & 600523 & & dailycaller.com & 771765 \\ 
 3 & & waynedupree.com & 258838 & & bongino.com & 346103 & & washingtonexaminer.com & 717017 \\ 
 4 & & judicialwatch.org & 233085 & & thenationalpulse.com & 215017 & & justthenews.com & 689725 \\ 
 5 & & truepundit.com & 176647 & & freebeacon.com & 197092 & & thefederalist.com & 687091 \\ 
 6 & & zerohedge.com & 165960 & & newsmax.com & 192924 & & dailywire.com & 396233 \\ 
 7 & & davidharrisjr.com & 150887 & & pjmedia.com & 123338 & & theepochtimes.com & 288656 \\ 
 8 & & politicalflare.com & 145838 & & newsbusters.org & 71008 & & nationalreview.com & 283172 \\ 
 9 & & djhjmedia.com & 112049 & & therightscoop.com & 66676 & & saraacarter.com & 267237 \\ 
 10 & & rumble.com & 101979 & & americanthinker.com & 59142 & & townhall.com & 256631 \\ 
 11 & & theconservativetreehouse.com & 99716 & &  &  & & theblaze.com & 191515 \\ 
 12 & & oann.com & 97325 & &  &  & & thepostmillennial.com & 181674 \\ 
 13 & & thedcpatriot.com & 90209 & & &  & & westernjournal.com & 165914 \\ 
 14 & & washingtonews.today & 79314 & &  &  & & redstate.com & 144010 \\ 
 15 & & rightwingtribune.com & 58442 & & &  & & thegreggjarrett.com & 139749 \\ 
 16 & & rt.com & 54985 & & &  & & bizpacreview.com & 97375 \\ 
 17 & & wnd.com & 54929 & & &  & & twitchy.com & 95401 \\ 
 18 & & gellerreport.com & 54277 & & &  & & trendingpolitics.com & 92094 \\ 
 19 & & nationalfile.com & 52393 & & &  & & lifenews.com & 90064 \\ 
 20 & & summit.news & 49539 & & &  & &  &  \\ 
 \\ \bottomrule
\end{tabular}
\end{adjustbox}
\vspace{0.1cm}

\begin{adjustbox}{width=0.65\columnwidth,center}
 \begin{tabular}{rllS[table-format = 6]llS[table-format = 6]llS[table-format = 6]}
\multicolumn{1}{l}{} &  & \multicolumn{2}{c}{right leaning news} &  & \multicolumn{2}{c}{center news} &  & \multicolumn{2}{c}{left leaning news} \\
                     &  & hostnames                   & $N$       &  & hostnames            & $N$       &  & hostnames              & $N$           \\ \toprule
1 & & nypost.com & 1701531 & & thehill.com & 2256888 & & nytimes.com & 6775402 \\ 
2 & & wsj.com & 887537 & & apnews.com & 1182504 & & washingtonpost.com & 6438506 \\ 
3 & & forbes.com & 748636 & & usatoday.com & 993957 & & cnn.com & 5577352 \\ 
4 & & washingtontimes.com & 408349 & & businessinsider.com & 773328 & & politico.com & 2290755 \\ 
5 & & foxbusiness.com & 212742 & & newsweek.com & 756820 & & nbcnews.com & 2231564 \\ 
6 & & thebulwark.com & 175417 & & reuters.com & 746033 & & theguardian.com & 1116515 \\ 
7 & & marketwatch.com & 96626 & & bbc.com & 296098 & & theatlantic.com & 1046475 \\ 
8 & & realclearpolitics.com & 93120 & & economist.com & 123939 & & abcnews.go.com & 1042419 \\ 
9 & & detroitnews.com & 77223 & & fivethirtyeight.com & 101824 & & npr.org & 871571 \\ 
10 & & dallasnews.com & 75910 & & ft.com & 91524 & & bloomberg.com & 767059 \\ 
11 & & rasmussenreports.com & 58712 & & foreignpolicy.com & 87729 & & cbsnews.com & 747442 \\ 
12 & & chicagotribune.com & 56974 & & factcheck.org & 79456 & & cnbc.com & 649041 \\ 
13 & & jpost.com & 55223 & & news.sky.com & 78372 & & axios.com & 621609 \\ 
14 & &  &  & &  &  & & msn.com & 613127 \\ 
15 & &  &  & &  &  & & news.yahoo.com & 586724 \\ 
16 & &  &  & &  &  & & independent.co.uk & 513765 \\ 
17 & &  &  & &  &  & & latimes.com & 451878 \\ 
18 & &  &  & &  &  & & citizensforethics.org & 382101 \\ 
19 & &  &  & &  &  & & buzzfeednews.com & 369962 \\      \\ \bottomrule
\end{tabular}
\end{adjustbox}

\vspace{0.1cm}
\begin{adjustbox}{width=0.5\columnwidth,center}
 \begin{tabular}{rllS[table-format = 6]llS[table-format = 6]}
\multicolumn{1}{l}{} &  & \multicolumn{2}{c}{left news}     &  & \multicolumn{2}{c}{extreme bias (left) news} \\
                     &  & hostnames              & $N$       &  & hostnames                   & $N$             \\ \toprule
1 & & rawstory.com & 2148200 & & occupydemocrats.com & 18151 \\ 
2 & & msnbc.com & 1606071 & & lancastercourier.com & 5815 \\ 
3 & & thedailybeast.com & 1404756 & & deepleftfield.info & 5753 \\ 
4 & & huffpost.com & 1121642 & & tplnews.com & 4022 \\ 
5 & & politicususa.com & 671043 & & bipartisanreport.com & 3243 \\ 
6 & & palmerreport.com & 434503 & & bossip.com & 2287 \\ 
7 & & motherjones.com & 424106 & & polipace.com & 586 \\ 
8 & & vox.com & 420613 & &  &  \\ 
9 & & vanityfair.com & 352964 & &  &  \\ 
10 & & nymag.com & 320049 & & &  \\ 
11 & & newyorker.com & 288409 & & &  \\ 
12 & & dailykos.com & 288384 & & &  \\ 
13 & & slate.com & 250942 & & &  \\ 
14 & & salon.com & 229583 & & &  \\ 
15 & & rollingstone.com & 190828 & & &  \\ 
16 & & thenation.com & 130272 & & &  \\ 
17 & & alternet.org & 126788 & & &  \\ 
18 & & theintercept.com & 104153 & & &   \\ \bottomrule
\end{tabular}
\end{adjustbox}
\caption{\textbf{Hostnames in each media category in 2020}.
We also show the number ($N$) of tweets 
with a URL pointing toward each hostname. 
Tweets with several URLs are counted multiple times.}
\label{tab:hostnames2020}
\end{table*}

\begin{table*}[!b]
\centering
\begin{adjustbox}{width=0.85\columnwidth,center}
\begin{tabular}{lS[table-format = 7]SS[table-format = 7]SSSSS}
\multicolumn{9}{c}{2016} \\
\toprule
                   & $N_t$                 & $p_t$                     & $N_u$                     & $p_u$                     & $N_t/N_u$                     & $p_{t,n/o}$                     & $p_{u,n/o}$                     & $N_{t,n/o}/N_{u,n/o}$                     \\
\midrule
 Fake news & 2991073 & 0.10 & 68391 & 0.03 & 43.73 & 0.19 & 0.07 & 124.22 \\ 
 Extreme bias right & 3969639 & 0.13 & 131346 & 0.06 & 30.22 & 0.09 & 0.05 & 56.73 \\ 
 Right news & 4032284 & 0.13 & 194229 & 0.08 & 20.76 & 0.11 & 0.07 & 33.77 \\ 
 Right leaning news & 1006746 & 0.03 & 64771 & 0.03 & 15.54 & 0.18 & 0.09 & 31.56 \\ 
 Center news & 6322257 & 0.21 & 600546 & 0.26 & 10.53 & 0.20 & 0.05 & 38.10 \\ 
 Left leaning news & 7491344 & 0.24 & 903689 & 0.39 & 8.29 & 0.14 & 0.06 & 19.16 \\ 
 Left news & 4353999 & 0.14 & 327411 & 0.14 & 13.30 & 0.14 & 0.07 & 26.16 \\ 
 Extreme bias left & 609503 & 0.02 & 19423 & 0.01 & 31.38 & 0.06 & 0.03 & 74.21 \\
\bottomrule
                   & \multicolumn{1}{l}{} & \multicolumn{1}{l}{}     & \multicolumn{1}{l}{}     & \multicolumn{1}{l}{}     & \multicolumn{1}{l}{}          & \multicolumn{1}{l}{}             & \multicolumn{1}{l}{}             & \multicolumn{1}{l}{}                          \\
\multicolumn{9}{c}{2020}                                        \\ 
\toprule
                   & $N_t$                 & \multicolumn{1}{c}{$p_t$} & \multicolumn{1}{c}{$N_u$} & \multicolumn{1}{c}{$p_u$} & \multicolumn{1}{c}{$N_t/N_u$} & \multicolumn{1}{c}{$p_{t,n/o}$} & \multicolumn{1}{c}{$p_{u,n/o}$} & \multicolumn{1}{c}{$N_{t,n/o}/N_{u,n/o}$} \\
\midrule
 Fake news & 4348747 & 0.06 & 99020 & 0.03 & 43.92 & 0.01 & 0.01 & 81.77 \\ 
 Extreme bias right & 4064820 & 0.06 & 107250 & 0.03 & 37.90 & 0.02 & 0.01 & 73.62 \\ 
 Right news & 8691901 & 0.12 & 382358 & 0.10 & 22.73 & 0.02 & 0.01 & 44.52 \\ 
 Right leaning news & 4648000 & 0.06 & 288207 & 0.08 & 16.13 & 0.02 & 0.01 & 23.35 \\ 
 Center news & 7568472 & 0.10 & 398241 & 0.11 & 19.00 & 0.03 & 0.02 & 33.96 \\ 
 Left leaning news & 33093267 & 0.45 & 2136830 & 0.59 & 15.49 & 0.03 & 0.02 & 22.85 \\ 
 Left news & 10513306 & 0.14 & 237685 & 0.07 & 44.23 & 0.03 & 0.02 & 73.42 \\ 
 Extreme bias left & 39857 & 0.00 & 887 & 0.00 & 44.93 & 0.05 & 0.02 & 82.59 \\ 
\bottomrule
\end{tabular}
\end{adjustbox}

\caption{\textbf{Tweet and user volume corresponding to each media category on Twitter between June $1^{\text{st}}$ until election day in 2016 (top) and 2020 (bottom)}. 
Number, $N_\textrm{t}$, and proportion, $p_\textrm{t}$, of tweets with a URL
  pointing to a website belonging to one of the media
  categories.  Number, $N_\textrm{u}$, and proportion, $p_\textrm{u}$, of unique users in each category. Users are classified in the category where the posted the largest number of tweets. Ties are randomly assigned.
  Proportion of tweets sent by
  non-official clients, $p_\textrm{t,n/o}$, proportion of users having sent
  at least one tweet from an non-official client, $p_\textrm{u,n/o}$, and
  average number of tweets per user sent from non-official clients,
  $N_\textrm{t,n/o}/N_{u,n/o}$.}
\label{tab:url_stats}
\end{table*}

\begin{table*}[ht!]
    \centering
    \begin{adjustbox}{width=0.85\columnwidth,center}
    \begin{tabular}{llrrrrrrr}
            & News category & Nodes & Edges & $\langle k \rangle$ & max($k_{out}$) & max($k_{in}$) & $\sigma(k_{out})/\langle k \rangle$ & $\sigma(k_{in})/\langle k \rangle$ \\ 
        \toprule
        \multirow{8}{*}{2016} & Fake News & 175,605 & 1,143,083 & 6.51 & 42,468 & 1232 & $32 \pm 4$ & $2.49 \pm 0.06$   \\
        & Extreme bias (right) & 249,659 & 1,637,927 & 6.56 & 51,845 & 588 & $36 \pm 6$ &  $2.73 \pm 0.03$  \\
        & Right & 345,644 & 1,797,023 & 5.20 & 86,454 & 490 & $44 \pm 11$ & $2.70 \pm 0.04$  \\
        & Right leaning & 216,026 & 495,307 & 2.29 & 32,653 & 129 & $45 \pm 11$ & $1.72 \pm 0.02$   \\
        & Center & 864,733 & 2,501,037 & 2.89 & 229,751 & 512 & $75 \pm 39$ & $2.69 \pm 0.06$  \\
        & Left leaning & 1,043,436 & 3,570,653 & 3.42 & 145,047 & 843 & $59 \pm 19$ & $3.38 \pm 0.10$ \\
        & Left & 536,903 & 1,801,658 & 3.36 & 58,901 & 733 & $47 \pm 12$ & $3.50 \pm 0.08$  \\
        & Extreme bias (left) & 78,911 & 277,483 & 3.52 & 23,168 & 648 & $33 \pm 6$ & $2.49 \pm 0.08$  \\ 
        \midrule
        \multirow{8}{*}{2020} & Fake News & 367,487 & 1,861,620 & 5.06 & 90,125 & 292 & $59 \pm 11$ & $2.05 \pm 0.02$   \\
        & Extreme bias (right) & 445,776 & 2,008,760 & 4.50 & 89,902 & 313 & $60 \pm 16$ & $2.09 \pm 0.02$  \\
        & Right & 674,935 & 4,452,861 & 6.59 & 109,053 & 607 & $54 \pm 9$ & $2.43 \pm 0.03$  \\
        & Right leaning & 882,552 & 3,203,999 & 3.63 & 115,302 & 298 & $59 \pm 16$ & $1.86 \pm 0.02$ \\
        & Center & 1,163,610 & 4,461,011 & 3.83 & 276,289 & 709 & $65 \pm 29$ & $2.37 \pm 0.04$ \\
        & Left leaning & 2,355,587 & 17,461,102 & 7.41 & 325,726 & 1,564 & $63 \pm 20$ & $3.62 \pm 0.05$ \\
        & Left & 819,684 & 4,688,119 & 5.71 & 175,841 & 1,042 & $57 \pm 14$ & $2.68 \pm 0.04$ \\
        & Extreme bias (left) & 21,411 & 26,888 & 1.25 & 5,755 & 27 & $41 \pm 3$ & $0.60 \pm 0.01$ \\
        \bottomrule
    \end{tabular}
    \end{adjustbox}
    \caption{Retweet network characteristics for each news category.
    Number of nodes, edges, average degree and degree heterogeneity of each network. The in- and out-degree heterogeneities are calculated by taking the average and standard error of 1000 independent samples of the degree heterogeneity ($\sigma(k_{in})/\langle k \rangle$ and $\sigma(k_{out})/\langle k \rangle$), each of which is computed on 78,911 samples with replacements from their respective degree distributions.}
    \label{tab:retweet_net}
\end{table*}

\begin{table*}[!htbp]
\begin{adjustbox}{width=0.7\columnwidth,center}
 \scriptsize
\centering
\begin{tabular}{rllll}
\toprule
{rank} &                 fake news                           &       extreme bias (right) news                                  &                         right news                    & right leaning news \\
       &  (7 verified, 2 deleted,                           &     (15 verified, 1 deleted,                           &              (22 verified, 0 deleted,                 &   (20 verified, 1 deleted                \\
       &                           19 unverified)             &                             9 unverified)              &                                      2 unverified)    &                           4 unverified)  \\
\midrule
1  &     {@PrisonPlanet}\,\checkmark &  {@realDonaldTrump}\,\checkmark &          {@FoxNews}\,\checkmark &              {@WSJ}\,\checkmark \\
2  &    {@RealAlexJones}\,\checkmark &      {@DailyCaller}\,\checkmark &  {@realDonaldTrump}\,\checkmark &        {@WashTimes}\,\checkmark \\
3  &                       {@zerohedge} &    {@BreitbartNews}\,\checkmark &       {@dcexaminer}\,\checkmark &           {@RT\_com}\,\checkmark \\
4  &                   {@DRUDGE\_REPORT} &        {@wikileaks}\,\checkmark &                   {@DRUDGE\_REPORT} &  {@realDonaldTrump}\,\checkmark \\
5  &  {@realDonaldTrump}\,\checkmark &                   {@DRUDGE\_REPORT} &           {@nypost}\,\checkmark &       {@RT\_America}\,\checkmark \\
6  &      {@mitchellvii}\,\checkmark &      {@seanhannity}\,\checkmark &   {@FoxNewsInsider}\,\checkmark &      {@WSJPolitics}\,\checkmark \\
7  &                         deleted                        &  {@WayneDupreeShow}\,\checkmark &        {@DailyMail}\,\checkmark &                   {@DRUDGE\_REPORT} \\
8  &                   {@TruthFeedNews} &                     {@LindaSuhler} &        {@AllenWest}\,\checkmark &   {@KellyannePolls}\,\checkmark \\
9  &                      {@RickRWells} &      {@mitchellvii}\,\checkmark &   {@RealJamesWoods}\,\checkmark &        {@TeamTrump}\,\checkmark \\
10 &                         deleted                        &         {@LouDobbs}\,\checkmark &    {@foxandfriends}\,\checkmark &         {@LouDobbs}\,\checkmark \\
11 &    {@gatewaypundit}\,\checkmark &     {@PrisonPlanet}\,\checkmark &        {@foxnation}\,\checkmark &  {@rebeccaballhaus}\,\checkmark \\
12 &                        {@infowars} &   {@DonaldJTrumpJr}\,\checkmark &         {@LouDobbs}\,\checkmark &       {@WSJopinion}\,\checkmark \\
13 &                   {@Lagartija\_Nix} &                  {@gerfingerpoken} &   {@KellyannePolls}\,\checkmark &      {@reidepstein}\,\checkmark \\
14 &   {@DonaldJTrumpJr}\,\checkmark &       {@FreeBeacon}\,\checkmark &    {@JudicialWatch}\,\checkmark &                         deleted                        \\
15 &                   {@ThePatriot143} &                 {@gerfingerpoken2} &     {@PrisonPlanet}\,\checkmark &  {@JasonMillerinDC}\,\checkmark \\
16 &                     {@V\_of\_Europe} &        {@TeamTrump}\,\checkmark &        {@wikileaks}\,\checkmark &       {@DanScavino}\,\checkmark \\
17 &                  {@KitDaniels1776} &                  {@Italians4Trump} &        {@TeamTrump}\,\checkmark &     {@PaulManafort}\,\checkmark \\
18 &                  {@Italians4Trump} &       {@benshapiro}\,\checkmark &    {@IngrahamAngle}\,\checkmark &         {@SopanDeb}\,\checkmark \\
19 &                        {@\_Makada\_} &   {@KellyannePolls}\,\checkmark &    {@marklevinshow}\,\checkmark &                      {@asamjulian} \\
20 &                    {@BigStick2013} &       {@DanScavino}\,\checkmark &        {@LifeZette}\,\checkmark &    {@JudicialWatch}\,\checkmark \\
21 &  {@conserv\_tribune}\,\checkmark &                         deleted                        &         {@theblaze}\,\checkmark &                        {@\_Makada\_} \\
22 &                     {@Miami4Trump} &                 {@JohnFromCranber} &      {@FoxBusiness}\,\checkmark &          {@mtracey}\,\checkmark \\
23 &                     {@MONAKatOILS} &                     {@true\_pundit} &  {@foxnewspolitics}\,\checkmark &                  {@Italians4Trump} \\
24 &                        {@JayS2629} &                   {@ThePatriot143} &                    {@BIZPACReview} &        {@Telegraph}\,\checkmark \\
25 &                      {@ARnews1936} &                        {@RealJack} &   {@DonaldJTrumpJr}\,\checkmark &    {@RealClearNews}\,\checkmark \\

\midrule
{rank} &           center news &                    left leaning news                                            &      left news  &                                     extreme bias (left) news \\
       &          (24 verified, 0 deleted, &        (25 verified, 0 deleted                                     &      (21 verified, 0 deleted, &                           (7 verified, 1 deleted, \\
       &           1 unverified)           &         0 unverified)                                              &      0 unverified) &                                   17 unverified)   \\
\midrule
1  &              {@CNN}\,\checkmark &         {@nytimes}\,\checkmark &       {@HuffPost}\,\checkmark &   {@Bipartisanism}\,\checkmark \\
2  &          {@thehill}\,\checkmark &  {@washingtonpost}\,\checkmark &           {@TIME}\,\checkmark &    {@PalmerReport}\,\checkmark \\
3  &         {@politico}\,\checkmark &             {@ABC}\,\checkmark &  {@thedailybeast}\,\checkmark &       {@peterdaou}\,\checkmark \\
4  &      {@CNNPolitics}\,\checkmark &         {@NBCNews}\,\checkmark &       {@RawStory}\,\checkmark &  {@crooksandliars}\,\checkmark \\
5  &          {@Reuters}\,\checkmark &           {@Slate}\,\checkmark &    {@HuffPostPol}\,\checkmark &                   {@BoldBlueWave} \\
6  &    {@NateSilver538}\,\checkmark &      {@PolitiFact}\,\checkmark &      {@NewYorker}\,\checkmark &       {@Shareblue}\,\checkmark \\
7  &               {@AP}\,\checkmark &         {@CBSNews}\,\checkmark &    {@MotherJones}\,\checkmark &                         {@Karoli} \\
8  &         {@business}\,\checkmark &       {@voxdotcom}\,\checkmark &            {@TPM}\,\checkmark &                  {@RealMuckmaker} \\
9  &         {@USATODAY}\,\checkmark &     {@ABCPolitics}\,\checkmark &          {@Salon}\,\checkmark &                   {@GinsburgJobs} \\
10 &      {@AP\_Politics}\,\checkmark &       {@ezraklein}\,\checkmark &  {@thinkprogress}\,\checkmark &                    {@AdamsFlaFan} \\
11 &  {@FiveThirtyEight}\,\checkmark &     {@nytpolitics}\,\checkmark &           {@mmfa}\,\checkmark &                       {@mcspocky} \\
12 &        {@bpolitics}\,\checkmark &        {@guardian}\,\checkmark &        {@joshtpm}\,\checkmark &    {@Shakestweetz}\,\checkmark \\
13 &       {@jaketapper}\,\checkmark &     {@NYDailyNews}\,\checkmark &          {@MSNBC}\,\checkmark &                        deleted                        \\
14 &                   {@DRUDGE\_REPORT} &         {@latimes}\,\checkmark &          {@NYMag}\,\checkmark &                        {@JSavoly} \\
15 &           {@cnnbrk}\,\checkmark &    {@BuzzFeedNews}\,\checkmark &       {@samstein}\,\checkmark &                {@OccupyDemocrats} \\
16 &  {@businessinsider}\,\checkmark &        {@Mediaite}\,\checkmark &      {@JuddLegum}\,\checkmark &                   {@ZaibatsuNews} \\
17 &            {@AC360}\,\checkmark &  {@HillaryClinton}\,\checkmark &       {@mashable}\,\checkmark &                       {@wvjoe911} \\
18 &             {@cnni}\,\checkmark &      {@nytopinion}\,\checkmark &   {@theintercept}\,\checkmark &    {@DebraMessing}\,\checkmark \\
19 &     {@brianstelter}\,\checkmark &     {@CillizzaCNN}\,\checkmark &    {@DavidCornDC}\,\checkmark &                     {@SayNoToGOP} \\
20 &   {@KellyannePolls}\,\checkmark &           {@MSNBC}\,\checkmark &       {@dailykos}\,\checkmark &                    {@coton\_luver} \\
21 &        {@wikileaks}\,\checkmark &           {@KFILE}\,\checkmark &     {@JoyAnnReid}\,\checkmark &                     {@EJLandwehr} \\
22 &         {@SopanDeb}\,\checkmark &     {@TheAtlantic}\,\checkmark &     {@nxthompson}\,\checkmark &                        {@mch7576} \\
23 &            {@KFILE}\,\checkmark &        {@SopanDeb}\,\checkmark &      {@thenation}\,\checkmark &                        {@RVAwonk} \\
24 &         {@BBCWorld}\,\checkmark &     {@Fahrenthold}\,\checkmark &      {@justinjm1}\,\checkmark &                         {@\_Carja} \\
25 &           {@NewDay}\,\checkmark &        {@BuzzFeed}\,\checkmark &    {@ariannahuff}\,\checkmark &                    {@Brasilmagic} \\
\bottomrule
\end{tabular}
\end{adjustbox}
\caption{\textbf{Top 25 CI {news spreaders} of the retweet networks corresponding to each media category in 2016}. 
Verified users have a checkmark (\checkmark) next to their username.
Verifying its accounts
is a feature offered by Twitter, that ``lets people know that an
account of public interest is
authentic''.
Unverified accounts do not have a checkmark and accounts marked as \textit{deleted} have been deleted, either by Twitter
or by the users themselves. Reproduced from \cite{Bovet2019}.
}
 \label{tab:influencers_2016}
\end{table*}

\begin{table*}[!htbp]
\begin{adjustbox}{width=0.7\columnwidth,center}
 \scriptsize
\centering
\begin{tabular}{rllll}
\toprule
{rank} &                 fake news                           &       extreme bias (right) news                                  &                         right news                    & right leaning news \\
       &  (10 verified, 8 deleted,                           &     (23 verified, 2 deleted,                           &              (23 verified, 1 deleted,                 &   (23 verified, 2 deleted                \\
       &                           7 unverified)             &                             0 unverified)              &                                      1 unverified)    &                           0 unverified)  \\
\midrule
    1 &                         @seanhannity$\checkmark$ &  (12) @DonaldJTrumpJr$\checkmark$ &  (25) @DonaldJTrumpJr$\checkmark$ &                                @nypost$\checkmark$ \\
    2 &                                          deleted &    (3) @BreitbartNews$\checkmark$ &   (19) @marklevinshow$\checkmark$ &              (1) @WSJ$\checkmark$ \\
    3 &                                  @DavidJHarrisJr &                              @dbongino$\checkmark$ &                                   @jsolomonReports &                        @DonaldJTrumpJr$\checkmark$ \\
    4 &                       @JudicialWatch$\checkmark$ &                         @marklevinshow$\checkmark$ &   (9) @RealJamesWoods$\checkmark$ &                             @EricTrump$\checkmark$ \\
    5 &                     @WayneDupreeShow$\checkmark$ &       (1) @realDonaldTrump &          (1) @FoxNews$\checkmark$ &       (4) @realDonaldTrump \\
    6 &                                        @catturd2 &                               @newsmax$\checkmark$ &                          @SaraCarterDC$\checkmark$ &        (2) @WashTimes$\checkmark$ \\
    7 &                           @TomFitton$\checkmark$ &                             @DailyMail$\checkmark$ &                           @DailyCaller$\checkmark$ &                         @marklevinshow$\checkmark$ \\
    8 &                                @OANN$\checkmark$ &                          @RaheemKassam$\checkmark$ &                           @MZHemingway$\checkmark$ &                              @brithume$\checkmark$ \\
    9 &                            @dbongino$\checkmark$ &                        @RealJamesWoods$\checkmark$ &                          @TrumpWarRoom$\checkmark$ &                        @RealJamesWoods$\checkmark$ \\
   10 &                                 @Thomas1774Paine &                            @joelpollak$\checkmark$ &       (3) @dcexaminer$\checkmark$ &                           @KimStrassel$\checkmark$ \\
   11 &                                   @RealMattCouch &                          @JackPosobiec$\checkmark$ &                          @JackPosobiec$\checkmark$ &                          @newtgingrich$\checkmark$ \\
   12 &                                          deleted &                             @TomFitton$\checkmark$ &                              @seanmdav$\checkmark$ &                          @TrumpWarRoom$\checkmark$ \\
   13 &                  (3) @zerohedge &                          @TrumpWarRoom$\checkmark$ &                         @realDailyWire$\checkmark$ &                                            deleted \\
   14 &                     @Rasmussen\_Poll$\checkmark$ &                          @RCamposDuffy$\checkmark$ &                         @GOPChairwoman$\checkmark$ &                        @MichaelCBender$\checkmark$ \\
   15 &                                        @atensnut &                             @EricTrump$\checkmark$ &       (2) @realDonaldTrump &                              @RandPaul$\checkmark$ \\
   16 &   (1) @PrisonPlanet$\checkmark$ &                       @JasonMillerinDC$\checkmark$ &                          @GreggJarrett$\checkmark$ &  (15) @JasonMillerinDC$\checkmark$ \\
   17 &                      @CassandraRules$\checkmark$ &      (14) @FreeBeacon$\checkmark$ &                          @newtgingrich$\checkmark$ &                          @JackPosobiec$\checkmark$ \\
   18 &                                          deleted &                            @AlexMarlow$\checkmark$ &                       @kayleighmcenany$\checkmark$ &                           @BillKristol$\checkmark$ \\
   19 &                        @DineshDSouza$\checkmark$ &                          @bennyjohnson$\checkmark$ &                        @RepDougCollins$\checkmark$ &                          @AriFleischer$\checkmark$ \\
   20 &     (5) @realDonaldTrump &                         @FrankelJeremy$\checkmark$ &                        @RichardGrenell$\checkmark$ &                       @Rasmussen\_Poll$\checkmark$ \\
   21 &                                  @HowleyReporter &                                            deleted &                       @AndrewCMcCarthy$\checkmark$ &                         @IngrahamAngle$\checkmark$ \\
   22 &                                          deleted &                            @SteveGuest$\checkmark$ &                            @SteveGuest$\checkmark$ &                          @RudyGiuliani$\checkmark$ \\
   23 &                                          deleted &                            @BrentScher$\checkmark$ &                         @SecretsBedard$\checkmark$ &                           @MZHemingway$\checkmark$ \\
   24 &                                          deleted &                         @IngrahamAngle$\checkmark$ &                              @parscale$\checkmark$ &                                @Forbes$\checkmark$ \\
   25 &                                          deleted &                          @kimguilfoyle$\checkmark$ &                              @dbongino$\checkmark$ &  (11) @rebeccaballhaus$\checkmark$ \\
\midrule
{rank} &           center news &                    left leaning news                                            &      left news  &                                     extreme bias (left) news \\
       &          (24 verified, 0 deleted, &        (24 verified, 0 deleted                                     &      (23 verified, 0 deleted, &                           (3 verified, 1 deleted, \\
       &           1 unverified)           &         1 unverified)                                              &      2 unverified) &                                   21 unverified)   \\
\midrule
    1 &          (2) @thehill$\checkmark$ &                                   @CNN$\checkmark$ &         (13) @MSNBC$\checkmark$ &                                    @DearAuntCrabby \\
    2 &               (7) @AP$\checkmark$ &          (1) @nytimes$\checkmark$ &  (3) @thedailybeast$\checkmark$ &                                @funder$\checkmark$ \\
    3 &          (5) @Reuters$\checkmark$ &                          @kylegriffin1$\checkmark$ &                        @kylegriffin1$\checkmark$ &                                   @ImpeachmentHour \\
    4 &                          @kylegriffin1$\checkmark$ &              (3) @ABC$\checkmark$ &   (19) @DavidCornDC$\checkmark$ &                                       @MeidasTouch \\
    5 &                             @JonLemire$\checkmark$ &   (2) @washingtonpost$\checkmark$ &       (1) @HuffPost$\checkmark$ &                       @TheDemCoalition$\checkmark$ \\
    6 &                              @Newsweek$\checkmark$ &                           @CNNPolitics$\checkmark$ &                       @NoahShachtman$\checkmark$ &                            @grantstern$\checkmark$ \\
    7 &                              @yarotrof$\checkmark$ &                                   @NPR$\checkmark$ &       (4) @RawStory$\checkmark$ &             (15) @OccupyDemocrats \\
    8 &         (9) @USATODAY$\checkmark$ &          (4) @NBCNews$\checkmark$ &    (7) @MotherJones$\checkmark$ &                                     @Stop\_Trump20 \\
    9 &                                    @ProjectLincoln &          (7) @CBSNews$\checkmark$ &                                      @TeaPainUSA &                                    @InSpiteOfTrump \\
   10 &                              @JoeBiden$\checkmark$ &                              @politico$\checkmark$ &                              @svdate$\checkmark$ &                                         @froggneal \\
   11 &                       @TheDemCoalition$\checkmark$ &                                @ddale8$\checkmark$ &                           @voxdotcom$\checkmark$ &                                            @atav1k \\
   12 &                          @TheEconomist$\checkmark$ &                              @CREWcrew$\checkmark$ &                              @maddow$\checkmark$ &                                     @diamondlilron \\
   13 &    (10) @AP\_Politics$\checkmark$ &                                @cnnbrk$\checkmark$ &                     @joncoopertweets$\checkmark$ &                                     @HollyHuntley3 \\
   14 &                         @TheRickWilson$\checkmark$ &                                @maddow$\checkmark$ &                               @Slate$\checkmark$ &                                            deleted \\
   15 &                              @tribelaw$\checkmark$ &                            @jaketapper$\checkmark$ &                      @PoliticusSarah$\checkmark$ &                                     @patrickinmass \\
   16 &                               @SkyNews$\checkmark$ &                         @ThePlumLineGS$\checkmark$ &                            @tribelaw$\checkmark$ &                                        @Franpianos \\
   17 &                                @maddow$\checkmark$ &                       @NatashaBertrand$\checkmark$ &                            @JoeBiden$\checkmark$ &                                        @willapercy \\
   18 &                        @FinancialTimes$\checkmark$ &                              @tribelaw$\checkmark$ &                       @TheRickWilson$\checkmark$ &                                        @Jerrygence \\
   19 &                       @joncoopertweets$\checkmark$ &                                 @axios$\checkmark$ &                                 @realTuckFrumper &                                         @bethlevin \\
   20 &                       @FrankFigliuzzi1$\checkmark$ &      (18) @nytopinion$\checkmark$ &                          @VanityFair$\checkmark$ &                                         @nyx\_with \\
   21 &                            @JimLaPorta$\checkmark$ &                             @maggieNYT$\checkmark$ &                            @CREWcrew$\checkmark$ &                                            @vg123e \\
   22 &                        @DonaldJTrumpJr$\checkmark$ &         (14) @latimes$\checkmark$ &      (6) @NewYorker$\checkmark$ &                                     @watercutter11 \\
   23 &        (24) @BBCWorld$\checkmark$ &                                    @ProjectLincoln &                        @PalmerReport$\checkmark$ &                                           @404HDTV \\
   24 &                           @APFactCheck$\checkmark$ &                             @60Minutes$\checkmark$ &                            @11thHour$\checkmark$ &                                        @j\_starace \\
   25 &                          @KamalaHarris$\checkmark$ &                              @business$\checkmark$ &    (5) @HuffPostPol$\checkmark$ &                                   @amberofmanyhats \\
\bottomrule
\end{tabular}
\end{adjustbox}
\caption{\textbf{Top 25 CI {news spreaders} of the retweet networks corresponding to each media category in 2020}. 
Verified users have a checkmark ($\checkmark$) next to their username. Unverified accounts do not have a checkmark and accounts marked as \textit{deleted} have been deleted, either by Twitter or by the users themselves. If a user held a position in the top 25 in 2016 as well, we mark that position for reference in parentheses next to the username. Despite \textit{@realDonaldTrump} having their account permanently suspended, due to the role they played in the 2020 Election, we have chosen to keep their original Twitter username in the table. However, we count this account as deleted, and have removed their previously assigned checkmark.
}
 \label{tab:influencers_2020}
\end{table*}

\newpage

\begin{table}[hp!]
\centering
\begin{tabular}{rrr|rr} 
\hline
\multicolumn{1}{l}{Year} & \multicolumn{1}{l}{Modularity (SE)} & \multicolumn{1}{l|}{Normalized Cut (SE)} & \multicolumn{1}{l}{Right Ratio} & \multicolumn{1}{l}{Left Ratio} \\ 
\hline
2016 & 0.234 (0.004) & 0.66 (0.03) & 0.038 & 0.05 \\
2020 & 0.236 (0.007) & 0.58 (0.03) & 0.038 & 0.08 \\
\hline
\end{tabular}
\caption{
    Tabulated analysis of the similarity network using quotes instead of retweets for the top influencers (as determined by the CI rankings of the retweet networks). The similarity network is found for the 2016 and 2020 data. Using Louvain community detection reveals two communities with left- and center-oriented influencers in one community, and right- and fake-oriented influencers in the other. Left side of table: average modularity and average normalized cut, with the standard errors (SE) in parentheses, determined by taking sub-samples of influencers from the quote similarity network, detecting the two dichotomous communities with the sub-sampled quote similarity network, then recording their modularities and normalized cuts. Right side of table: ratio of quotes-to-retweets within the complete similarity network. Specifically, number of user quotes of influencer tweets over number of user retweets of influencer tweets. Right ratio indicates the average ratio for the community with right-oriented influencers. Left ratio indicates the average ratio for the community with left-oriented influencers. These ratios are found for both 2016 and 2020.}
    \label{tab:quote_similarity}
\end{table}

\FloatBarrier

\begin{table}[hp!]
\begin{center}

\begin{minipage}{0.5\linewidth}
    \begin{tabular}{rrllc}
    & & \multicolumn{3}{c}{\textbf{A} overall quotes/retweets}      \\           
    & \multicolumn{1}{l|}{} & \multicolumn{1}{c}{2016} &  & \multicolumn{1}{c}{2020} \\ \cline{2-3} \cline{5-5}
    \multirow{2}{*}{from users} & \multicolumn{1}{r|}{right}                 & 0.03                   &  & 0.03                   \\
     & \multicolumn{1}{r|}{left}                  & 0.05                   &  & 0.04                  
    \end{tabular}
\end{minipage}    

\begin{minipage}{0.5\linewidth}
    \begin{tabular}{rllllll}
    & & \multicolumn{5}{c}{\textbf{B} quotes/retweets}      \\         
    &                           & \multicolumn{2}{c}{2016} &  & \multicolumn{2}{c}{2020} \\
    &                           & \multicolumn{5}{c}{to influencers} \\                           
    & \multicolumn{1}{l|}{}      & right       & left       &  & right       & left       \\ \cline{2-4} \cline{6-7} 
 \multirow{2}{*}{from users}    & \multicolumn{1}{r|}{right} & 0.02      & 0.19     &  & 0.02      & 0.49     \\
    & \multicolumn{1}{r|}{left}  & 0.56      & 0.03     &  & 3.76      & 0.03    
    \end{tabular}
\end{minipage}
    \begin{minipage}{0.5\linewidth}

\end{minipage}
\end{center}
\caption{Comparison of fraction of retweets and quotes from users to influencers with different latent ideology estimates.
Users and influencers are divided in two categories based on their ideology estimates, namely left (ideology $<$0) and right (ideology$>$0). Table \textbf{A} shows the overall proportion of quotes over retweets from users on the right and on the left revealing that the number of quotes represent only a small fraction ($\leq 5\%$) of the number of retweets. Table \textbf{B} shows the proportion of quotes over retweets from users to influencers for all pairs of ideology categories in 2016 and in 2020.}
\label{tab:rt_vs_quote}
\end{table}

\newpage

\begin{table}[hp!]
\begin{adjustbox}{width=\columnwidth,center}
\centering
\begin{tabular}{@{}rlllllllllll@{}}
                         & \multicolumn{5}{c}{users distributions}                                                            & \multicolumn{1}{c}{} & \multicolumn{5}{c}{influencers distributions}                                                      \\
                             & \multicolumn{1}{c}{2016} & \multicolumn{1}{c}{95\% CI} &         \multicolumn{1}{c}{2020} & \multicolumn{1}{c}{95\% CI} & \multicolumn{1}{c}{difference} & \multicolumn{1}{c}{} & \multicolumn{1}{c}{2016} & \multicolumn{1}{c}{95\% CI} &  \multicolumn{1}{c}{2020} & \multicolumn{1}{c}{95\% CI} &  \multicolumn{1}{c}{difference} \\ \cmidrule(lr){2-6} \cmidrule(l){8-12} 
all                          & \multicolumn{1}{c}{0.1086} &
\multicolumn{1}{c}{[0.1082,0.1091]}&
\multicolumn{1}{c}{0.1474} &
\multicolumn{1}{c}{[0.1471,0.1477]}&
\multicolumn{1}{c}{0.0388}       & \multicolumn{1}{c}{} & \multicolumn{1}{c}{0.1786} &
\multicolumn{1}{c}{[0.1606,0.1965]}&
\multicolumn{1}{c}{0.2091} &
\multicolumn{1}{c}{[0.1907,0.2282}&
\multicolumn{1}{c}{0.0305}       \\
common users                 & \multicolumn{1}{c}{0.0941} &
\multicolumn{1}{c}{[0.0934,0.0947]}&
\multicolumn{1}{c}{0.1172} &
\multicolumn{1}{c}{[0.1166,0.1178]}&
\multicolumn{1}{c}{0.0231}       & \multicolumn{1}{c}{} & \multicolumn{1}{c}{0.1793} &
\multicolumn{1}{c}{[0.1616,0.1979]}&
\multicolumn{1}{c}{0.2143} &
\multicolumn{1}{c}{[0.1952,0.2336]}&
\multicolumn{1}{c}{0.0350}       \\
common influencers           & \multicolumn{1}{c}{0.1070} &
\multicolumn{1}{c}{[0.1065,0.1076]}&
\multicolumn{1}{c}{0.1830} &
\multicolumn{1}{c}{[0.1825,0.1834]}&
\multicolumn{1}{c}{0.0760}       & \multicolumn{1}{c}{} & \multicolumn{1}{c}{0.1641} &
\multicolumn{1}{c}{[0.1290,0.1951]}&
\multicolumn{1}{c}{0.1741} &
\multicolumn{1}{c}{[0.1376,0.2122]}&
\multicolumn{1}{c}{0.0100}       \\
\begin{tabular}[c]{@{}r@{}}common users\\ and influencers\end{tabular} & \multicolumn{1}{c}{0.0947} &
\multicolumn{1}{c}{[0.0940,0.0955]}&
\multicolumn{1}{c}{0.1399} &
\multicolumn{1}{c}{[0.1390,0.1406]}&
\multicolumn{1}{c}{0.0452}       & \multicolumn{1}{c}{} & \multicolumn{1}{c}{0.1650} &
\multicolumn{1}{c}{[0.1314,0.2034]}&
\multicolumn{1}{c}{0.1719} &
\multicolumn{1}{c}{[0.1379,0.2086]}&
\multicolumn{1}{c}{0.0069}       \\
\end{tabular}
\end{adjustbox}
\caption{Hartigans' dip test statistics of the users and influencers latent ideology distributions when considering all users and influencers, only users that were present in 2016 and 2020, only influencers that were present in 2016 and 2020 and only users and influencers that were present in 2016 and 2020.
95\% confidence intervals are computed from 1000 bootstrap samples with the bias-corrected and accelerated confidence intervals method.}
\label{tab:dip_tests}
\end{table}

\newpage
\FloatBarrier

\bibliography{supplement}
\bibliographystyle{ieeetr}